\documentclass[fleqn,usenatbib]{mnras}
\usepackage{newtxtext,newtxmath}

\usepackage[T1]{fontenc}

\DeclareRobustCommand{\VAN}[3]{#2}
\let\VANthebibliography\thebibliography
\def\thebibliography{\DeclareRobustCommand{\VAN}[3]{##3}\VANthebibliography}

\usepackage{graphicx}	
\usepackage{amsmath}	
\usepackage{threeparttable}

\graphicspath{{./}{figures_ito/}}

\title[A QG with a Type 1 AGN at $z=2.09$]{Black hole mass of a quiescent galaxy hosting a Type 1 AGN at z=2.09: Implications for black hole - galaxy co-evolution and AGN quenching at high redshift}

\author[K. Ito et al.]{
Kei Ito,$^{1,2,3}$\thanks{E-mail: kei.ito.astro@gmail.com (KI)}
Takumi S. Tanaka,$^{1,4,5}$
Kazuhiro Shimasaku,$^{1,6}$
Makoto Ando,$^{7}$
Masafusa Onoue,$^{4,5,8}$
\newauthor
Masayuki Tanaka,$^{7,9}$
Suin Matsui,$^{1}$
Takumi Kakimoto,$^{7,9}$
Francesco Valentino$^{2,3,10}$
\\
$^{1}$Department of Astronomy, Graduate School of Science, The University of Tokyo, 7-3-1 Hongo, Bunkyo-ku, Tokyo 113-0033, Japan\\
$^{2}$ Cosmic Dawn Center (DAWN), Denmark \\
$^{3}$ DTU Space, Technical University of Denmark, Elektrovej 327, DK2800 Kgs. Lyngby, Denmark \\
$^{4}$Kavli Institute for the Physics and Mathematics of the Universe, The University of Tokyo, Kashiwa, Chiba 277-8583, Japan\\
$^{5}$Center for Data-Driven Discovery, Kavli IPMU (WPI), UTIAS, The University of Tokyo, Kashiwa, Chiba 277-8583, Japan\\
$^{6}$Research Center for the Early Universe, Graduate School of Science, The University of Tokyo, 7-3-1 Hongo, Bunkyo-ku, Tokyo 113-0033, Japan\\
$^{7}$National Astronomical Observatory of Japan, 2-21-1 Osawa, Mitaka, Tokyo, 181-8588, Japan\\
$^{8}$Kavli Institute for Astronomy and Astrophysics, Peking University, Beijing 100871, China\\
$^{9}$Department of Astronomical Science, The Graduate University for Advanced Studies, SOKENDAI, 2-21-1 Osawa, Mitaka, Tokyo, 181-8588, Japan\\
$^{10}$European Southern Observatory, Karl-Schwarzschild-Str. 2, D-85748 Garching bei Munchen, Germany\\
}

\date{Accepted XXX. Received YYY; in original form ZZZ}

\pubyear{2024}

\begin{document}
\label{firstpage}
\pagerange{\pageref{firstpage}--\pageref{lastpage}}
\maketitle

\begin{abstract}
We report a characterization of an X-ray-detected quiescent galaxy at $z=2.09$, named COS-XQG1, using JWST/NIRCam and NIRSpec data. This galaxy is detected in Chandra imaging, suggesting the presence of an AGN with a high black hole accretion rate of $\dot{M}_{\rm BH}=0.22\pm0.03\, {\rm M_\odot yr^{-1}}$. Using multi-wavelength photometry from X-ray to sub-millimeter, including the latest JWST imaging, we confirm that COS-XQG1 is massive ($M_\star = (1.6\pm0.2)\times10^{11}\, M_\odot$) and quiescent (${\rm sSFR}<10^{-10}\, {\rm yr^{-1}}$) as reported previously, even considering the contribution from AGN emission. Noticeably, COS-XQG1 displays a broad H$\beta$ and H$\alpha$ emission component with a full width at half maximum of $4365^{+81}_{-81}\, {\rm km\, s^{-1}}$ in its NIRSpec spectrum, which is typical of Type 1 AGNs. The line width and luminosity of the broad H$\alpha$ emission give a black hole mass of $\log{(M_{\rm BH}/M_\odot)} = 8.43\pm0.02\, (\pm 0.5)$. With a stellar velocity dispersion measurement ($\sigma_\star=238\pm32\, {\rm km\, s^{-1}}$), we find that this galaxy is consistent with the local relations in the $M_{\rm BH} - \sigma_\star$ and $M_{\rm BH}- M_\star$ planes, which might suggest that massive quiescent galaxies at $z\geq2$ have already been mature in terms of both stellar and black hole masses and will not evolve significantly. In addition, image 2D-decomposition analysis finds that this galaxy comprises disk and point source components. The latter is likely the composition of an AGN and a stellar bulge. Based on a comparison with numerical simulations, we expect that COS-XQG1 will evolve into a typical quiescent galaxy with lower AGN activity by redshift 0. This study shows the usefulness of X-ray-detected quiescent galaxies in investigating the co-evolution between SMBHs and galaxies in the early Universe.
\end{abstract}




\section{Introduction}

The relation between black holes and galaxy properties has been intensively investigated for decades. At $z\sim0$, there are tight correlations between black hole mass and galaxy properties such as stellar velocity dispersion ($\sigma_\star$) of the bulge component, bulge mass, or total stellar mass \citep[e.g.,][]{Magorrian1998, Haring2004, Bennert2011, Kormendy2013, Reines2015, Greene:2020:ARA&A}. These tight correlations indicate a connection between the growth of galaxies and black holes. Active galactic nuclei (AGNs) at high redshift have been used to understand the origin of this co-evolution \citep[e.g.,][]{Willott:2017:ApJ, Maiolino:2023:arXiv}. Recently, the James Webb Space Telescope (JWST) confirmed the existence of numerous populations of low-mass AGNs at $z\sim5-10$ \citep[e.g.,][]{Kocevski:2023:ApJL, Kocevski:2024:arXiv, Harikane:2023:ApJ, Maiolino2023_GNz11, Maiolino:2023:arXiv}. 

According to the current models and simulations, AGNs are one of the causes of galaxy quenching. Their radiation, wind, or jet can prevent gas inside galaxies from transforming into stars \citep[see][for a review]{Fabian:2012:ARA&A}. In this sense, progenitors of local quenched galaxies at high redshift can provide insights into the early connection between AGNs and quenching. Thanks to deep imaging and spectroscopy, it is now known that some massive galaxies have little star formation activity compared to typical star-forming galaxies, even at high redshift. These `quiescent galaxies' are now found at $z\sim4-5$ \citep{Tanaka:2019:ApJL, Kakimoto:2024:ApJ, Barrufet:2024:arXiv, Antwi-Danso:2023:arXiv, Carnall:2023:Natur, Carnall:2024:arXiv, deGraaff:2024:arXiv}. It is reported that quiescent galaxies at high redshift have high X-ray and radio luminosities likely from AGNs \citep[e.g.,][]{Olsen:2013:ApJ, Ito:2022:ApJ}, and several quiescent galaxies have emission lines whose flux ratio suggests that they are likely to originate from AGNs \citep[e.g.,][]{Kriek:2009:ApJL, Belli:2019:ApJ, Belli:2023:arXiv, Nanayakkara:2024:NatSR, deGraaff:2024:arXiv, Park:2024:arXiv}. However, the properties of the AGNs in quiescent galaxies at high redshift are poorly understood. 

The black hole mass of quiescent galaxies at high redshift is a critical property that provides insights into the co-evolution between black holes and galaxies and the connection between AGN and quenching. The spectra of quiescent galaxies often have strong absorption lines in the rest-frame optical wavelength; thus some studies successfully constrain the stellar velocity dispersion even at high redshift \citep[e.g.,][]{Tanaka:2019:ApJL, Esdaile:2021:ApJL, Forrest:2022:ApJ}. 
If quiescent galaxies with $\sigma_\star$ measurements also have black hole mass measurements, we can use them to examine the massive end of the $M_{\rm BH} - \sigma_\star$ relation. In addition, cosmological hydrodynamical simulations show that black hole mass is the most predictive parameter of whether galaxies are quenched or not \citep[e.g.,][]{Bluck:2023:ApJ}. At $z>2$, however, only one quiescent galaxy has a black hole mass measurement. This galaxy is located at $z=4.66$, and its black hole mass has been obtained from the broad H$\alpha$ emission \citep{Carnall:2023:Natur}.

We need to increase the sample of quiescent galaxies with clear AGN features to investigate the co-evolution of black holes and galaxies and the connection between AGN and quenching. A plausible method to construct such a sample is to use X-ray emission. Suppose a galaxy at high redshift (e.g., $z\geq2$) is detected in a current deep X-ray image, such as taken with the Chandra telescope.
In that case, its X-ray emission is likely from the AGN since such high emission cannot be explained by other sources, e.g., X-ray binaries, unless it is a starburst galaxy. Previous spectroscopic campaigns on high-redshift quiescent galaxies already confirmed those with significant X-ray emission \citep[e.g.,][]{Belli:2014:ApJL, Belli:2017:ApJ, D'Eugenio:2021:A&A}. Infrared spectroscopy of these galaxies enables us to constrain the black hole mass based on their broad emission lines if they are Type 1 AGNs. 

This paper reports a characterization of one X-ray-detected quiescent galaxy at $z=2.09$, observed with JWST NIRCam and NIRSpec, called COS-XQG1. This galaxy was first reported in \citet{Belli:2014:ApJL}\footnote{The ID of this galaxy in \citet{Belli:2014:ApJL} is 31769.}. We derive its black hole mass from its broad H$\alpha$ emission, the second case for constraining the black hole mass of quiescent galaxies at $z>2$. Also, through 2D decomposition image analysis and SED fitting, we investigate host galaxy properties, especially its morphology. 

This paper is organized as follows. In section \ref{sec:data}, we describe the available data of COS-XQG1 and derive the stellar velocity dispersion and global properties of this galaxy, such as star formation rate and stellar mass. The black hole mass and accretion rate are derived in Section \ref{sec:BH}. We perform 2D decomposition analysis in Section \ref{sec:gal}, where its morphology is investigated. We plot COS-XQG1 in the $M_{\rm BH} - \sigma_\star$ and $M_{\rm BH} - M_\star$ relations and discuss the black hole-galaxy co-evolution in Section 5. Our results are summarized in Section \ref{sec:concl}. We assume a $\Lambda$CDM cosmology with $H_0=70\ {\rm km\ s^{-1}Mpc^{-1}}$, $\Omega_m = 0.3$, and $\Omega_\Lambda = 0.7$. Magnitudes are based on the AB magnitude system \citep{Oke:1983:ApJ}.  

\section{Target and Data}\label{sec:data}
\subsection{Target: COS-XQG1}
\par The target of this paper, COS-XQG1, which is known as XID-159 and CID-960 in the XMM-Newton COSMOS X-Ray Point Source Catalog \citep{Cappelluti2009} and the Chandra-COSMOS Legacy Survey Point Source Catalog \citep{Civano:2016:ApJ}, respectively, is a galaxy located in the COSMOS field. It has a spectroscopic redshift of $z=2.096$ by Keck/MOSFIRE \citep{Belli:2014:ApJL} \footnote{We have updated its redshift to $z=2.0943 $ using a NIRSpec spectrum (see Section \ref{subsec:ppxf}).}. Its colors in the $UVJ$-diagram \citep{Williams:2009:ApJ} and its absorption features, such as Ca{\sc ii} lines, seen in its Keck/MOSFIRE $J$-band spectrum, suggest that this galaxy is in the quiescent phase. \citet{Yuan:2014:ApJL} report that this galaxy is likely in a rich cluster at $z=2.095$.  
\par Two features suggest that this galaxy has an AGN. One is that this galaxy is detected in the Chandra COSMOS Legacy Survey \citep{Civano:2016:ApJ} with signal-to-noise ratios of 7.26, 7.01, and 3.14 in the 0.5-7\, keV, 0.5-2\, keV, and 2-7\, keV, respectively. Its observed X-ray luminosity in the rest-frame 2-10 keV calculated from the soft band flux is $(8.9\pm1.3)\times10^{43}\, {\rm erg/s}$ in the assumption of photon index as $\Gamma=1.8$. Its hardness ratio, $HR = (H-S)/(H+S)$, where $H$ and $S$ are the count rates in the hard and soft bands, respectively, is $-0.44\pm0.14$. This low value implies that the AGN does not have significant obscuration. The other feature is a broad H$\alpha$ emission line in a MOSFIRE $K$-band spectrum taken in part of the ZFIRE survey \citep{Nanayakkara:2016:ApJ} (see Appendix \ref{sec:MOSFIREHalpha}). This feature is discussed in detail with its JWST/NIRSpec spectrum in Section \ref{subsec:halpha}. 

\subsection{Imaging data}\label{subsec:imaging_data}
PRIMER (Public Release IMaging for Extragalactic Research) is a $\sim$200-hours JWST Cycle 1 GO Treasury imaging survey (GO1837, PI. James Dunlop).
PRIMER observation covers two HST CANDELS Legacy Fields, COSMOS and UDS, utilizing eight filters in JWST/NIRCam (\citealt{Rieke2023}; F090W, F115W, F150W, F200W, F277W, F356W, F410M, F444W) and two filters in JWST/MIRI (\citealt{Bouchet2015}; F770W and F1800W).

COS-XQG1 is covered by PRIMER observations targeting the COSMOS field. In addition to the PRIMER NIRCam and MIRI observations, we also use HST/ACS F606W \citep{Koekemoer2011} and F814W \citep{Koekemoer2007} images. In this study, we use imaging data processed and made available through the DAWN {\it JWST} Archive (DJA\footnote{\url{https://dawn-cph.github.io/dja/}}). 
The image reduction is conducted using {\sc Grizli} \footnote{\url{https://github.com/gbrammer/grizli}} (see \citealt{Valentino2023} for details of the image reduction process), and the pixel scale is 0\farcs04/pixel.

\subsection{Spectroscopic data}
\par COS-XQG1 was observed with JWST/NIRSpec in the Cycle1 GO Program 1810 (PI. Sirio Belli). The observation was conducted in the MSA mode with the G140M, G235M, and G395M ($R\sim1000$) grism. We use the first two spectra. See \citet{Davies:2024:MNRAS} for more details of the survey design and observation. The existence of the Na{\sc id} absorption and broad emission lines is briefly reported in that paper\footnote{The ID of COS-XQG1 is 12020 in \citet{Davies:2024:MNRAS}.}. 
\par In this paper, we use the spectra compiled in 
 the spectroscopic part of DJA. The reduction of the spectrum in DJA is performed using the {\sc Grizli} and {\sc MSAExp} \footnote{\url{https://github.com/gbrammer/msaexp}} software. An 1D spectrum is obtained from the reduced 2D spectrum using the optimal extraction method \citep{Horne:1986:PASP}. For more details on the reduction and extraction procedures, refer to \citet{Heintz:2024:arXiv}. 
 
\par
\begin{figure}
    \centering
    \includegraphics[width=8cm]{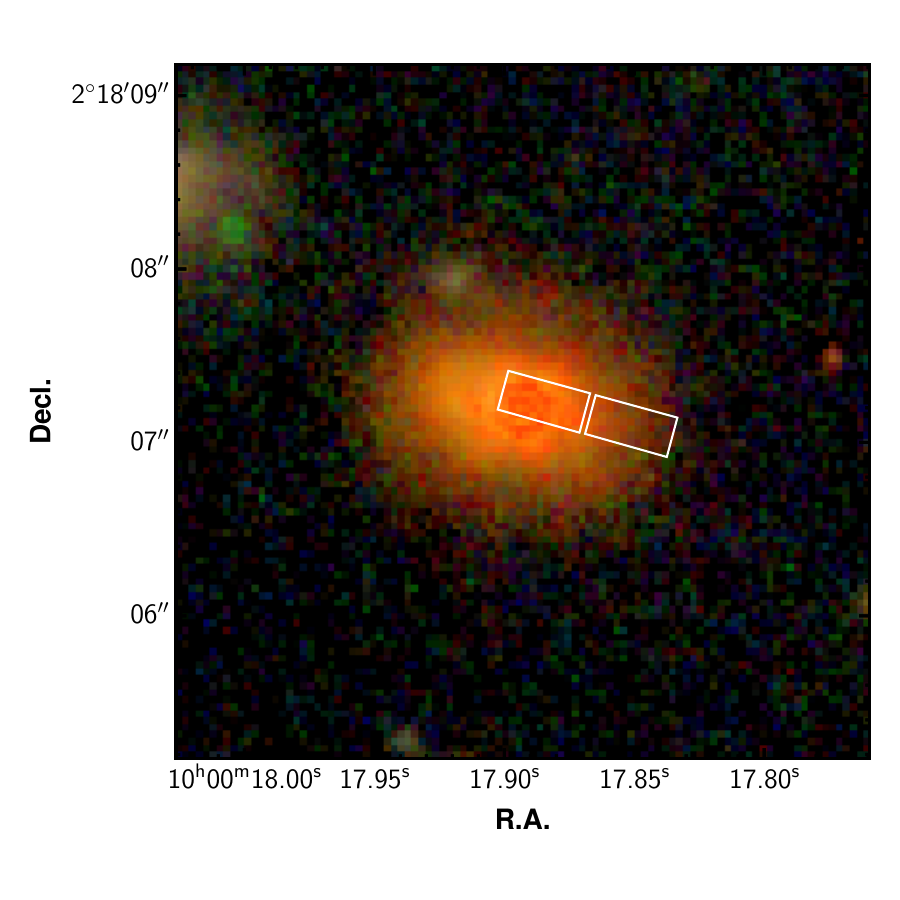}
    \caption{Three-color JWST/NIRCam image of COS-XQG1, with the shutter position in the JWST/NIRSpec observation shown by white polygons. F090W, F150W, and F200W images are assigned to blue, green, and red, respectively.}\label{fig:slitpos}
\end{figure}

\begin{figure*}
    \centering
    \includegraphics[width=17cm]{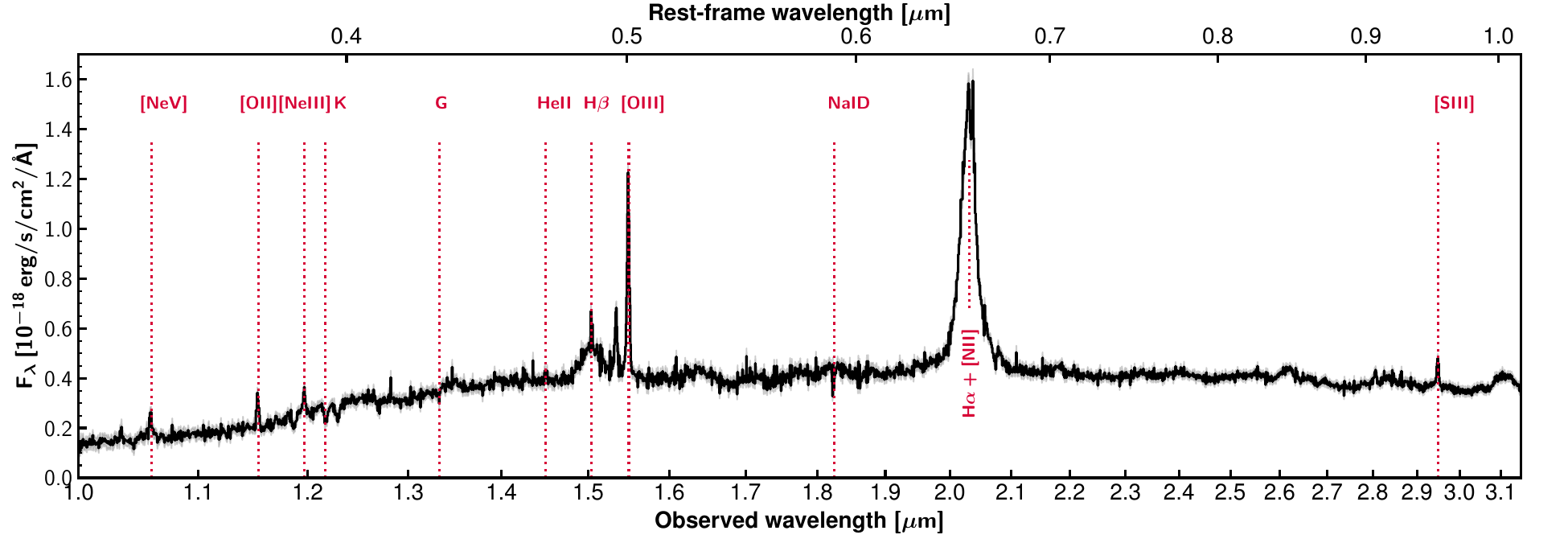}
    \caption{The JWST/NIRSpec spectrum of COS-XQG1. The black line is the spectrum, and the surrounding gray-shaded region corresponds to its $1\sigma$ uncertainty. Major spectral features are highlighted.}
    \label{fig:spec-all}
\end{figure*}

 Further flux calibration is needed to avoid the mismatch of the spectral shapes to that of SED models due to imperfect flux calibration and slit-loss correction. We adopt a 2nd-order polynomial correction function to match the spectrum with the photometry of {\it JWST}/NIRCam whose filters' wavelengths lie within the wavelength range of the NIRSpec spectrum. We choose photometry measured within $0.5\arcsec$ diameter since the ratio between the flux within $0.5\arcsec$ diameter and the total flux is not strongly dependent on wavelength for point sources, which is $\sim 0.85$\, (0.80 - 0.88 across the wavelength range in question). The variation of this ratio is likely due to the different PSF sizes. This trend means that the photometry traces the same SED shape as the intrinsic one. We do not attempt to calibrate the spectrum based on the total photometry because the MSA shutter is open only at the galaxy's center (Figure \ref{fig:slitpos}). 
This choice has a negligible impact on the AGN property from the spectrum (e.g., black hole mass) because the AGN should be a point source. 
We will find that this galaxy consists of a central compact component and a disk (Section \ref{subsec:compact}), which may have different stellar populations. The observed spectrum, taken only from the central region, might be dominated by the bulge and is inadequate to calibrate using the total flux. Therefore, we do not derive global properties of this galaxy, e.g., stellar mass, from the spectrum. Figure \ref{fig:spec-all} shows the NIRSpec spectrum used in this study. 
\subsection{SED fitting analysis with global photometry}\label{subsec:CIGALE}
We perform an SED fitting analysis using {\sc XCIGALE} \citep{Yang:2020:MNRAS}. This analysis aims to obtain global physical properties of COS-XQG1, such as stellar mass and star formation rate. Since COS-XQG1 is detected in X-ray, AGN emission is expected to contribute to its SED shape. In this sense, {\sc XCIGALE} is an adequate tool since it contains AGN templates. We use its version 2022.1 \citep{Yang:2022:ApJ}.

Not only the photometry from JWST but also the photometry by other telescopes available in the COSMOS field is used in the fitting. We here use the optical to IR photometry compiled as the COSMOS2020 catalog \citep{Weaver:2022:ApJS} and the super-deblended far-infrared to (sub) millimeter catalog from \citet{Jin:2018:ApJ}. Note that there is no bright source around COS-XQG1, at least in the rest-frame UV/optical imaging; thus, we expect a negligible impact from the nearby sources in the photometry. Specifically, we use the following band data from the COSMOS2020 Classic Catalog: $U$ band from CFHT/MegaCam, $grizy$ bands and some medium/narrow bands from Subaru/HSC and Suprime-Cam, $YJHKs$ bands from VISTA/VIRCAM, and channels 1,2,3 and 4 from Spitzer/IRAC. From the super-debldended catalog, we use the data of Spitzer/MIPS 24\,$\mu$m, Hershel/PACS 100\,$\mu$m, 160\,$\mu$m, and JCMT/SCUBA2 850\,$\mu$m. The X-ray fluxes from the Chandra COSMOS Legacy Survey \citep{Civano:2016:ApJ} are also used. We do not include the photometry of HST/F606W and F814W since the HSC imaging covers the same wavelength range with similar sensitivities. In total, we use 45 data points. For the JWST/NIRCam data, we use the total photometry corrected from 0.5\arcsec aperture photometry of JWST/NIRCam imaging summarized in DJA. For the photometry in the COSMOS2020 catalog, the total photometry corrected from 2\arcsec aperture photometry is employed. The total photometry of JWST/MIRI imaging is derived by summing up the pixel values. The Milky Way dust extinction is corrected using the dust map from \citet{Schlafly:2011:ApJ}.

In the fitting, we include a stellar continuum, nebular emission, galactic dust emission, AGN emission, and dust attenuation. We assume a delayed star formation history (SFH) expressed as:
\begin{equation}
    {\rm SFR}(t) \propto \frac{t}{\tau^2}\exp{(-t/\tau)}\, {\rm for}\, 0\leq t \leq t_0
\end{equation}
where $\tau$ is the e-folding time, $t_0$ is the age of the onset of the star formation, and $t$ is the look-back time. We vary $\tau$ over $0.1\, {\rm Gyr}< \tau < 3\, {\rm Gyr}$, and $0.001\, {\rm Gyr}$ (equivalent to a single burst). The stellar age is set to be between $0.05\, {\rm Gyr}$ and $3.5\, {\rm Gyr}$ (the cosmic age at $z=2.09$). We use the stellar synthesis model from \citet{Bruzual:2003:MNRAS}, assuming a \citet{Chabrier:2003:PASP} initial mass function and a Solar metallicity ($Z=Z_\odot$). Nebular emission is calculated using \citet{Inoue:2011:MNRAS} with an ionization parameter of $ \log{U}= -3$. Dust attenuation is implemented based on \citet{Calzetti:2000:ApJ} with a color excess of $0\leq E (B-V)\leq 2$. Galactic dust emission is implemented using an IR spectral template from \citet{Dale2014}. Lastly, the SED of the AGN component is modeled with {\tt SKIRTOR} clumpy torus model \citep{Stalevski:2012:MNRAS, Stalevski:2016:MNRAS}, with an SED shape of the AGN accretion disk taken from \citet{Schartmann:2005:A&A}. Among the AGN parameters, the viewing angle, power-law index modifying the optical slope of the disk \citep[see Equation 5 in][]{Yang:2022:ApJ}, the AGN fraction, and the extinction in the polar direction are set as free parameters. Their parameter ranges are $0\deg<i<90\deg$, $-1\leq \delta<1$, $0\leq f_{\rm AGN}\leq0.999$, and $0\leq E (B-V)\leq0.8$, respectively. The photon index of the X-ray spectrum is fixed as $\Gamma=1.8$. The X-ray emission is linked to the AGN continuum via $\alpha_{OX}$, which is defined as $\alpha_{OX} = 0.3838\log{(L_{\nu\, 2{\rm keV}}/L_{\nu, 0.25{\rm \mu m}})}$, where $L_{\nu\, 2{\rm keV}}$ and $L_{\nu, 0.25{\rm \mu m}}$ are the luminosity density at $2\, {\rm keV}$ and $0.25\, {\rm \mu m}$, respectively. {\sc XCIGALE} imposes that this $\alpha_{OX}$ should follow the empirical relation from \citet{Just:2007:ApJ} with a deviation less than $\Delta \alpha_{OX}<1$.

\begin{figure*}
    \centering
    \includegraphics[width=17cm]{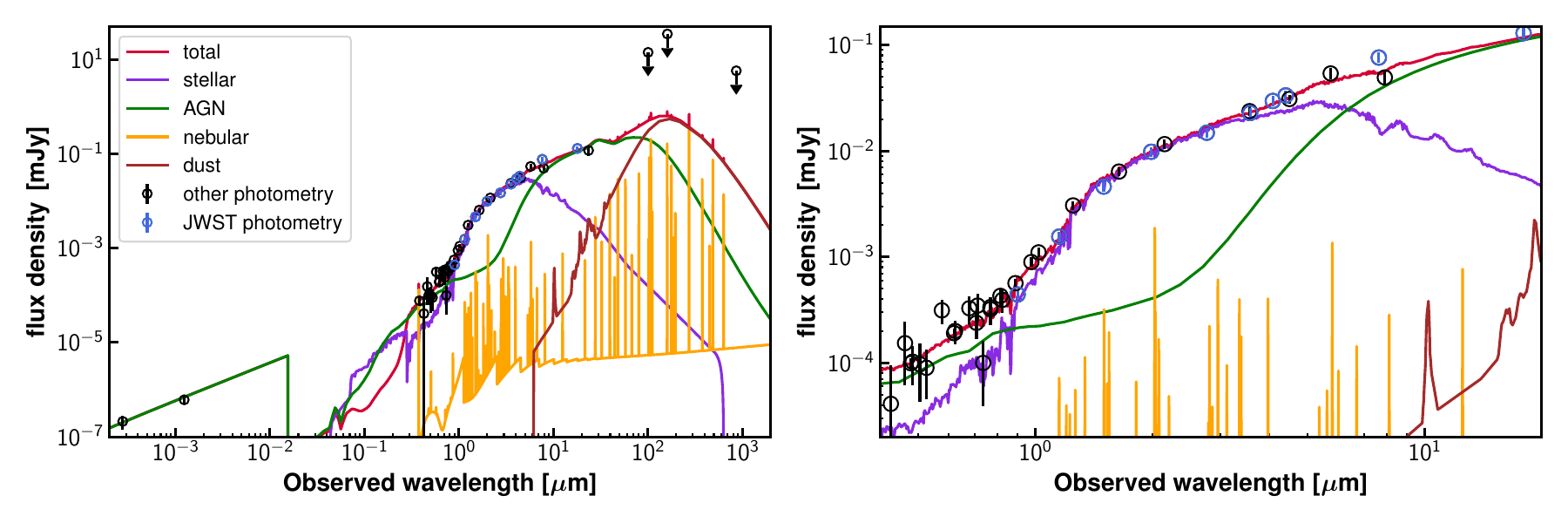}
    \caption{Summary of {\sc XCIGALE} fitting for COS-XQG1. The left panel shows the best-fit SED in the entire wavelength range, and the right panel shows the zoom-in view for $\lambda_{\rm obs}=0.4-20\, {\rm \mu m}$. Blue circles show the JWST photometry, while black ones show the photometry from other telescopes. The fluxes not detected with the signal-to-noise ratio more than three are shown as $3\sigma$ upper limits. Red, purple, green, yellow, and brown lines are the best-fit SED of the total stellar continuum, AGN, nebular, and dust emission, respectively.}
    \label{fig:CIGALE-SED}
\end{figure*}

Figure \ref{fig:CIGALE-SED} shows the best fit of this {\sc XCIGALE} fitting. The reduced chi-square of this fitting is $\chi^2_\nu=1.0$. Its stellar mass is estimated as $M_\star = (1.6 \pm 0.2) \times 10^{11}\, M_\odot$. The star formation rate (SFR) is constrained to be ${\rm SFR} < 12.8 M_\odot\, {\rm yr^{-1}}$ as the $2\sigma$ upper limit with the best-fit value as $1.5\, M_\odot\, {\rm yr^{-1}}$. Here, the instantaneous SFR is used. This leads to the specific star formation rate as ${\rm sSFR} < 10^{-10}\, {\rm yr^{-1}}$ with the best-fit value as ${\rm sSFR}=0.9\times10^{-11}\, {\rm yr^{-1}}$.  This sSFR is more than one dex lower than that of the star formation main sequence \citep[e.g.,][]{Schreiber:2015:A&A}, suggesting the quiescent nature of COS-XQG1 even if we consider the contribution of the AGN component to the SED. Physical properties of COS-XQG1, including those from other analyses, are summarised in Table \ref{tab:XQG1-summary}.

The best-fit SED in Figure \ref{fig:CIGALE-SED} shows that the AGN continuum is negligible compared to the stellar continuum at $1\, {\rm \mu m}<\lambda_{\rm obs}<10\, {\rm \mu m}$. The AGN continuum flux density at rest-frame $2500$ \AA\, and $\alpha_{OX}$ calculated from the best-fit SED are $(10.4\pm0.7)\times10^{29}\, {\rm erg\, s^{-1}\, Hz^{-1}}$ and $\alpha_{OX}=-1.5\pm0.03$, respectively. These values are in agreement with those of QSOs over a wide redshift range of $0<z<6$ \citep[e.g.,][]{Just:2007:ApJ, Nanni:2017:A&A} and are consistent with the scaling relation between the rest-frame $2500$ \AA\, AGN continuum luminosity and $\alpha_{OX}$ \citep[e.g.,][]{Just:2007:ApJ} within the $1\sigma$ uncertainty. In other words, such a negligible contribution of the AGN continuum to the total SED shape in the rest-frame optical is as expected from the QSO continuum from other observations.

Following the previous studies \citep[e.g.,][]{Schreiber:2018:A&A, Belli:2019:ApJ, Estrada-Carpenter:2020:ApJ, D'Eugenio:2021:A&A, Valentino:2020:ApJ, Carnall:2023:Natur}, we define the stellar age as the look-back time after 50\% of the stellar mass was formed. This $t_{\rm 50}$ is derived from the best-fit values and their uncertainties of $t_0$ and $\tau$. We find that COS-XQG1 was formed $1.3$ Gyrs before the observed epoch.

\begin{table}
\scriptsize
\caption{Physical Properties of COS-XQG1.}\label{tab:XQG1-summary}
\begin{threeparttable}[t]
    \centering
    \begin{tabular}{ccc} 
    \hline
    Quantity & Symbol & Value\\
    \hline
    R.A. (J2000) & - & 10:00:17.90 \\
    Decl. (J2000) & - & +02:18:07.20 \\ 
    Redshift & $z$ & $2.0943 \pm 0.0003$ \\ 
    Stellar mass & $M_\star/M_\odot$ & $(1.6 \pm 0.2) \times 10^{11}$\\
    Star formation rate from SED fitting & ${\rm SFR_{SED}}/(M_\odot\, {\rm yr^{-1}})$ & $<12.8$\tnote{a} \\
    Star formation rate from narrow H$\alpha$ emission & ${\rm SFR_{H\alpha}}/(M_\odot\, {\rm yr^{-1}})$ & $1.5\pm0.2$\tnote{b} \\
    Dust attenuation & $E(B-V)/{\rm mag}$ & $0.21\pm0.07$ \\
    Stellar age & $t_{\rm 50}/{\rm Gyr}$ & $1.34 \pm 0.52$ \\ 
    Stellar velocity dispersion & $\sigma_\star/({\rm km\, s^{-1}})$ & $238 \pm 32$ \\
    Black hole mass & $\log{(M_{\rm BH}/M_\odot)}$ & $8.43\pm0.02\ (\pm0.5)$ \\
    Bolometric luminosity from X-ray & $L_{\rm X,\, bol}/({\rm erg\, s^{-1}})$ & $(1.4\pm0.2)\times10^{45}$ \\
    Bolometric luminosity from H$\alpha$ & $L_{\rm H\alpha,\, bol}/({\rm erg\, s^{-1}})$ & $(2.18\pm0.05)\times10^{45}$ \\
    Black hole accretion rate & $\dot{M}_{\rm BH}/(M_\odot\, {\rm yr^{-1}})$ & $0.22\pm0.03$ \\
    Eddington ratio & $ L_{\rm X,\, bol}/L_{\rm edd}$ & $0.037^{+0.080}_{-0.025}$\\
    \hline
    \end{tabular}
    \begin{tablenotes}
    \item[a] {This is the $2\sigma$ upper limit of the SFR derived from the SED fitting with {\sc XCIGALE}.}
    \item[b] {This is estimated by assuming the narrow H$\alpha$ emission line is only from the star formation.}
    \end{tablenotes}
\end{threeparttable}
\end{table}

\subsection{Stellar velocity dispersion measurement}\label{subsec:ppxf}

Stellar velocity dispersion is measured by fitting the NIRSpec spectrum using {\sc ppxf} \citep{Cappellari:2017:MNRAS, Cappellari:2023:MNRAS}. We use stellar templates generated from the flexible stellar population synthesis model \citep{Conroy:2009:ApJ}, where templates older than the cosmic age of the redshift of the object are removed. Furthermore, we limit to templates with metallicities of $\log{(Z/Z_\odot)}\geq 0$. In the fitting, we include gaseous emission lines, i.e., Balmer emission lines, [O{\sc ii}]$\lambda\lambda$3726,\, 3729, [Ne{\sc iii}]$\lambda\lambda$3968,\, 3869, He{\sc ii}$\lambda$4687, [O{\sc iii}]$\lambda$5007, [O{\sc i}]$\lambda$6300,  He{\sc i}$\lambda$5876, [N{\sc ii}]$\lambda\lambda6548,\, 6584$, and [S{\sc ii}]$\lambda\lambda6716,\, 6731$, which are described as Gaussian functions. We also have another Gaussian component in the wavelength H$\alpha$, H$\beta$, and H$\gamma$ to consider the broad line component. The velocity center is fixed to the value of the narrow lines. More detailed line fitting for broad lines is conducted in Section \ref{subsec:halpha}. We use a 2nd order multiplicative correction function and no additive correction in the fitting, following \citet{Cappellari:2023:MNRAS}. The stellar templates have a spectral resolution of $\sim2.5$\,\AA\ FWHM at $3540$\,\AA$<\lambda<$ $7350$\,\AA; thus we fit the spectrum in this rest-frame wavelength range. We use the full wavelength range to obtain the best fit at a wavelength around H$\alpha$, which is used in Section \ref{subsec:halpha}, but even if we only use the spectrum at the rest-frame wavelength of $3540$\,\AA$<\lambda<4200$\,\AA, the results will not change significantly. The fitting is conducted twice. After the first fitting, we mask the spectral pixels having $>3\sigma$ higher/lower values than the best fit and fit again, as suggested in \citet{Cappellari:2023:MNRAS}.

\begin{figure*}
    \centering
    \includegraphics[width=17cm]{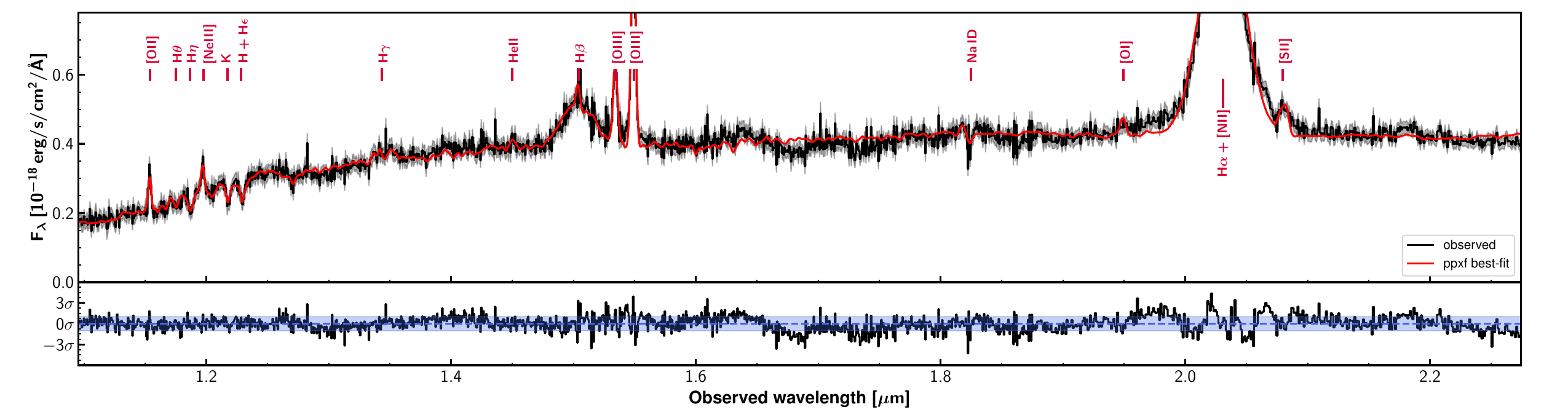}
    \caption{Summary of {\sc ppxf} fitting for COS-XQG1. Top panel: The red line corresponds to the best-fit spectrum model from {\sc ppxf}, whereas the black lines and hatched region correspond to the observed spectrum and its $1\sigma$ uncertainty. Bottom panel: The difference between the observed spectrum and the best-fit model in the unit of the observed uncertainty. The blue hatched region corresponds to $1\sigma$.}
    \label{fig:spec-ppxf}
\end{figure*}

The best-fit spectrum is shown in Figure \ref{fig:spec-ppxf}. The fit was well with a reduced chi-square of $\chi_\nu^2=0.83$. We obtain the stellar velocity dispersion as $\sigma_\star = 238 \pm 32 \, {\rm km\, s^{-1}}$ after correcting the instrumental dispersion. Its uncertainty is derived by a Monte-Carlo simulation, in which we perturb the spectrum using noise spectra and perform the fitting 100 times. We employ the 68 percentile of the distribution of stellar velocity dispersion as its uncertainty. In addition, we obtain a slightly revised redshift of $z=2.0943 \pm 0.0003$ from that reported in \citet{Belli:2014:ApJL} in this fitting, which is used in this study.

\section{Black hole Properties}\label{sec:BH}
\subsection{Line fitting}\label{subsec:halpha}
\par As also seen in the MOSFIRE spectrum, a significantly broad H$\alpha$ emission line is detected in the NIRSpec spectrum at $\lambda_{\rm obs}\sim2.03\, {\rm \mu m}$ (Figure \ref{fig:spec-all}). This emission line has two peaks, which can be due to the blending with [N{\sc ii}] doublet emission lines. In addition, the H$\beta$ line also has broad emission. Here, we simultaneously fit the spectrum around the H$\alpha$ and H$\beta$ lines. The main aim of this fitting is to derive the flux and line width of the broad H$\alpha$ emission line, with which we will constrain the black hole mass. Note that the broad H$\beta$ and H$\alpha$ emission are included in the {\sc ppxf} fitting assuming it follows the Gaussian function (Section \ref{subsec:ppxf}), but a more detailed fitting is needed because a significant residual is seen at the outskirts of the broad H$\alpha$ emission in the best-fitting (see Figure \ref{fig:spec-ppxf}).
\par We assume that the Voigt profile describes the broad component of H$\beta$ and H$\alpha$ emission. The narrow component of H$\beta$ and H$\alpha$ and the other lines ([O{\sc iii}], [N{\sc ii}] and [S{\sc ii}]) is assumed to be a Gaussian profile. We assume no velocity offsets between these lines. In addition, all narrow components are assumed to have the same line width. The line flux ratios of the [N{\sc ii}]\ and [O{\sc iii}]\ doublet were fixed to 1:2.94 and 1:3, respectively \citep[e.g.,][]{Osterbrock:2006:agna}. We thus have 12 free parameters in the fitting, i.e., velocity center, broad H$\alpha$ flux, broad H$\beta$ flux, narrow H$\alpha$ flux, narrow H$\beta$ flux, [O{\sc iii}] flux, [N{\sc ii}] flux, [S{\sc ii}]$\lambda6716$ flux, [S{\sc ii}]$\lambda6731$ flux, full width at half maximum (FWHM) of the Lorentzian part of the Voigt profile for the broad component, FWHM of the Gaussian part of the Voigt profile for the broad component, and the FWHM of the narrow Gaussian component. We use the continuum subtracted spectrum to model the emission lines. The best fit from the spectral modeling in Section \ref{subsec:ppxf} is employed as the continuum spectrum. Using the best-fit model as a continuum spectrum enables us to deal with not only the continuum slope but also a possible H$\alpha$ absorption line in the host galaxy spectrum, which is likely to exist according to the significant absorption lines at $\lambda_{\rm rest}\sim0.4\, {\rm \mu m}$ (see Figure \ref{fig:spec-ppxf}). We use the flux within $\pm 10000\, {\rm km s^{-1}}$ and $\pm 15000\, {\rm km s^{-1}}$ around the H$\beta$ and H$\alpha$ emission lines to fit the line profile. The fitting is conducted using the Markov Chain Monte Carlo method. 

\par In Figure \ref{fig:Halpha}, we show the fitting result. The posterior distribution of each parameter is summarized in Appendix \ref{app:conerplot}. The reduced chi-square of this fitting is $\chi^2_\nu=1.1$. The total flux and the FWHM of the broad H$\alpha$ emission line are estimated as $f_{\rm broad\ H\alpha}=(
3.86^{+0.05}_{-0.05})\times10^{-16}\, {\rm erg\, s^{-1}\, cm^{-2}}$ and ${\rm FWHM_{H\alpha}}=4365^{+81}_{-81}\, {\rm km\, s^{-1}}$, respectively. These values clearly show the existence of a broad H$\alpha$ emission. The large FWHM suggests that this is from the broad line region. The [N{\sc ii}]$\lambda6585$ emission line is also significantly detected with a flux of $f_{\rm [N{\sc II}]\lambda 6585}=(8.4\pm 1.3)\times 10^{-18}\, {\rm erg\, s^{-1}\, cm^{-2}}$. 

\par Both line ratio [O{\sc iii}]${\rm \lambda5007/H\beta}$ and [N{\sc ii}]${\rm \lambda6585/H\alpha}$ ratios are high, which are $\log$[O{\sc iii}]$\lambda5007/{\rm H\beta} = 0.75^{+0.05}_{-0.05}$ and $\log$[N{\sc ii}]$\lambda6585/{\rm H\alpha} = -0.03^{+0.07}_{-0.07}$, respectively. These high line ratios imply that these emission lines are mainly from the AGN activity \citep[e.g.,][]{ Kauffmann:2003:MNRAS, Kewley:2013:ApJ}.

\par This line fitting enables us to estimate the SFR in an independent way from the SED fitting. We convert the observed H$\alpha$ narrow line flux to the SFR following \citet{Kennicutt:1998:ARA&A} adjusted for our choice of a \citet{Chabrier:2003:PASP} IMF. Ignoring the dust extinction because of the high line ratio between H$\alpha$ and H$\beta$ (${\rm H\beta/H\alpha} = 1.8^{+0.4}_{-0.3}$), we derive the SFRs as ${\rm SFR} = 1.5 \pm 0.2\, M_\odot\, {\rm yr^{-1}}$. This value is perfectly consistent with that from the SED fitting with {\sc XCIGALE}. Thus, this line fitting also supports the quiescent nature of the COS-XQG1. Noted that this SFR from the narrow H$\alpha$ emission line should be treated as an upper limit since we assume no contribution in the H$\alpha$ emission line from the central AGN.

\par Some of recent studies which constrain the broad H$\alpha$ emission use the Gaussian profile \citep[e.g.,][]{Matthee:2024:ApJ, Maiolino:2023:arXiv} or the Lorentzian profile \citep[e.g.,][]{Kollatschny:2013:A&A, Loiacono:2024:A&A}. We try fitting using the Gaussian function instead of the Voigt function, finding that the Gaussian function provides a worse fitting result, with a reduced chi-square of the fitting of $\chi^2_{\nu}=2$. We also find that using the Lorentzian function gives an FWHM consistent with that from the Voigt profile. We also try including additional broad Gaussian components as a possible outflow component in addition to the narrow Gaussian components and Voigt components. This is motivated by some potential residuals seen around [O{\sc iii}]$\lambda5007$ (see Figure \ref{fig:Halpha}). We find that including these additional broad components hardly changes the fluxes of the original broad and narrow components and hence the SFR and black hole mass measurements and the line diagnostic result.
\begin{figure*}
    \includegraphics[width=17cm]{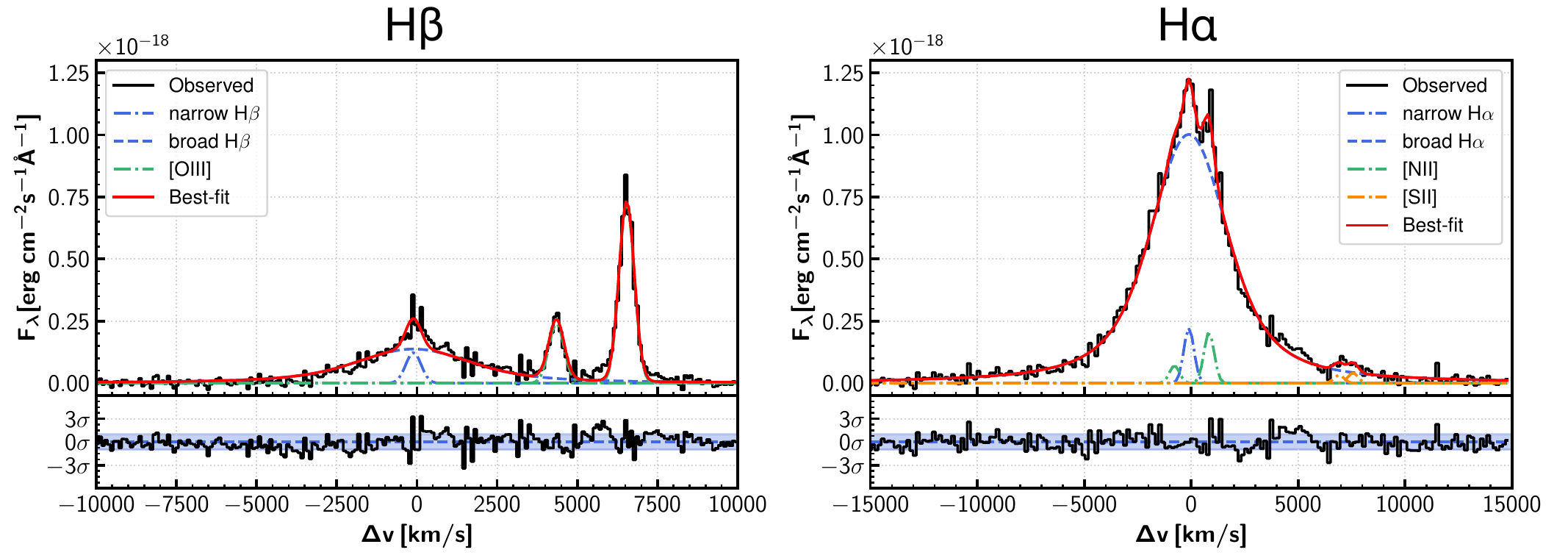}
    \caption{Left panel: Spectral fitting around H$\beta$ line. The upper panel shows the observed spectrum (black), best-fit (red), narrow H$\beta$ in the best-fit (blue dashed-dotted), broad H$\beta$ in the best-fit (blue dashed), and [O{\sc iii}] in the best-fit (green dashed-dotted light). The lower panel shows the fitting residual normalized by the spectrum's uncertainty. Right panel: Spectral fitting around H$\alpha$ line. The meaning of the upper and lower panels is the same as the left panel. In the upper panel, the narrow H$\alpha$ in the best-fit (blue dashed-dotted), broad H$\alpha$ in the best-fit (blue dashed), [N{\sc ii}] in the best-fit (green dashed-dotted light), and [S{\sc ii}] in the best-fit (orange dashed-dotted) are shown.}
    \label{fig:Halpha}
\end{figure*}

\subsection{Black hole mass and Eddington Ratio} \label{subsec:bhmass}
The broad H$\alpha$ line profile, high X-ray luminosity, and properties of the other lines suggest that the broad line is from the AGN activity. Thus, we derive the black hole mass using the measured flux and FWHM of the broad H$\alpha$ emission, as done in the other studies on broad H$\alpha$ lines at high redshift \citep[e.g.,][]{Carnall:2023:Natur, Harikane:2023:ApJ, Ubler:2023:A&A, Maiolino:2023:arXiv, Matthee:2024:ApJ}. We utilized the scaling relation from \citet{Reines:2013:ApJ}:
\begin{equation}
\begin{split}
    \log{(M_{\rm BH}/M_\odot)} = 6.57+\log{(\epsilon)}+0.47\log{(L_{\rm H\alpha,broad}/10^{42}\, {\rm erg\, s^{-1}})}\\
    +2.06\log{({\rm FWHM}_{\rm H\alpha,broad}/10^3\, {\rm km\, s^{-1}})}, \label{eq:BHmass}
\end{split}
\end{equation}
where $\epsilon$ is the geometric correction factor and we assume $\epsilon=1.075$, following \citet{Reines2015}. The $L_{\rm H\alpha,broad}$ and ${\rm FWHM}_{\rm H\alpha,broad}$ are the luminosity and FWHM of the broad H$\alpha$ emission, respectively. Using the best-fit parameters obtained in Section \ref{subsec:halpha}, we obtain the black hole mass of COS-XQG1 as:
\begin{equation}
\log{(M_{\rm BH}/M_\odot)} = 8.43\pm0.02.
\end{equation}
It should be noted that there is a systematic uncertainty in black hole masses derived from Equation \ref{eq:BHmass}, which is $\sim0.5$ dex according to \citet{Shen2013}. 
\par Since X-ray emission is detected for this galaxy, we derive the bolometric luminosity and the black hole accretion rate from its X-ray luminosity. Using the conversion factor $k_{\rm bol}$ ($k_{\rm bol} \equiv L_{\rm bol}/L_X$) given in \citet{Yang:2018:MNRAS}, we derive a bolometric luminosity of $L_{\rm bol} = (1.4\pm0.2)\times10^{45}\, {\rm erg\, s^{-1}}$. 
The contribution from X-ray binaries to $L_X$ is negligible according to an empirical scaling relation that predicts the total $L_X$ of X-ray binaries in a galaxy from its stellar mass and SFR \citep[e.g.,][]{Fornasini:2018:ApJ}. Thus, we do not correct for it. We note that the $L_{\rm bol}$ obtained above broadly agrees with the bolometric luminosity derived by converting the broad H$\alpha$ luminosity to the AGN continuum luminosity using the relation between the rest-frame 5100\AA\ luminosity and the H$\alpha$ luminosity \citep{Greene:2005:ApJ} and the relation between the rest-frame 5100\AA\ luminosity and the bolometric luminosity \citep{Richards:2006:ApJS}, as done in \citet{Harikane:2023:ApJ} ($L_{\rm bol} = (2.18\pm0.05)\times10^{45}\, {\rm erg\, s^{-1}}$). The black hole accretion rate (BHAR) is then derived from the bolometric luminosity as follows:
\begin{equation}
    \dot{M}_{\rm BH}= \frac{(1-\epsilon)\times L_{\rm bol}}{\epsilon c^2},
\end{equation}
where $c$ is the speed of light and $\epsilon$ is the efficiency of the mass conversion into radiation. We assume $\epsilon=0.1$. As a result, we obtain the BHAR as 
\begin{equation}
    \dot{M}_{\rm BH}= 0.22\pm0.03\, M_\odot\, {\rm yr^{-1}}.
\end{equation}

The Eddington luminosity is related to the black hole mass as:
\begin{equation}
    L_{\rm edd} = 1.3\times10^{38}(M_{\rm BH}/M_\odot)\, {\rm erg\, s^{-1}}.
\end{equation}
Using the estimated black hole mass, the Eddington luminosity of COS-XQG1 is estimated as $L_{\rm edd} = 3.7^{+7.9}_{-2.5}\times 10^{46}\, {\rm erg\, s^{-1}}$, where the error includes the systematic uncertainty on the black hole mass measurement. Thus, the Eddington ratio of COS-XQG1 is estimated as $L_{\rm bol}/L_{\rm edd} = 0.037^{+0.08}_{-0.025}$. The physical parameters of the black hole obtained in this subsection are also summarized in Table \ref{tab:XQG1-summary}. 

\section{Host-Galaxy Properties}\label{sec:gal}

\subsection{2D decomposition analysis}\label{subsec:decomposition}
We can reconstruct the image of the host galaxies by subtracting the AGN image from the original image utilizing 2D decomposition analysis, where we fit the original image by the composition of the PSF and a Sérsic profile \citep{sersic1968}. After the launch of JWST, thanks to its high spatial resolution and sensitivity, 2D decomposition analysis yields higher quality host-only images at higher $z$ \citep{Ding2022_CEERS, Ding2022_z6, Li2023, Kocevski2023, Zhang2023, Harikane:2023:ApJ, Zhuang2023COSMOS, Tanaka2024}.
We can detect substructures of host galaxies even at $z\sim2$ \citep{Zhuang2023COSMOS, Tanaka2024}. 

\begin{figure}
    \includegraphics[width=8.5cm]{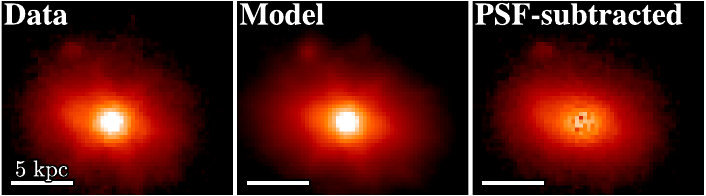}
    \caption{Example 2D decomposition fitting in F277W.
    (Left) Original image.
    (Middle) Model image, which is composed of a PSF component and Sérsic components.
    (Right) PSF subtracted image, i.e., Original $-$ PSF model.
    The host-only image is successfully extracted. The size of each image is $2\farcs4\times2\farcs0$, and the lower left white bars in individual panels indicate 5~kpc (corresponding to $\sim0\farcs6$).
    The full filter version is shown as Figure~\ref{fig:decom_all} in Appendix~\ref{app:deom_full}.}
    \label{fig:decom}
\end{figure}

We perform a 2D decomposition analysis of HST+JWST images (Section~\ref{subsec:imaging_data}) using the 2D image analysis tool {\sc GaLight} \citep{Ding2020_HST}. 
With {\sc GaLight}, we conduct forward modeling of the original image as a superposition of a PSF component and a PSF-convolved Sérsic component. 
We use the same position for the PSF component with a small buffer of $\pm$0\farcs04, corresponding to $\pm1$ pixel, among the fitted filters.
The center position of the Sérsic component is free from the PSF component, i.e., their alignment is not necessarily required as suggested in \cite{Yue2023}.
To avoid unrealistic results, we limit the parameter range of the effective radius ($r_e$) and the Sérsic index ($n$) within [0\farcs08, 2\farcs0] ($\sim$[0.50~kpc, 13~kpc]) and [0.3, 7], respectively.
Finally, based on the fitting results, we extract host-only images by subtracting the PSF component from the original image, as shown in Figure~\ref{fig:decom}.

In 2D decomposition analysis, accurate subtraction of the AGN-derived PSF component is crucial, requiring PSF reconstruction \citep[e.g.,][]{Zhuang2023, Tanaka2024}.
This study follows the PSF reconstruction strategy by \cite{Tanaka2024}.
We construct a PSF library from star images in the same mosaic image using {\tt find\_PSF} function in {\sc GaLight}.
Then, we perform a fitting of COS-XQG1 using each star image as a PSF, and the star images with the lowest $\chi^2_\nu$ values are stacked to create the final PSF (Top-5 PSF).
Note that, similar to \cite{Tanaka2024}, we confirm that results do not significantly change when using a Top-75\% PSF, which is a stack of the PSFs with the lowest 75\% $\chi^2_\nu$ values, or a PSF modeled with {\sc PSFEx} \citep{Bertin2011}. 
Previous studies have reported that errors associated with PSF uncertainty are much larger than fitting errors in 2D decomposition analysis from JWST images \citep{Yue2023, Ding2022_z6, Tanaka2024}.
Therefore, in this study, we assess the error by $\chi^2_\nu$-based weighting of each single PSF result, following the approach of \cite{Ding2020_HST}.
The estimated $r_e$, $n$, and the magnitude of the PSF and the Sérsic components in each filter are summarized in Table~\ref{tab:molphology}. Due to the combination of high spatial resolution (i.e., small PSF size) and high $S/N$, the estimated parameters show the smallest errors in F200W.

\begin{table}
\scriptsize
\caption{Summary of fitted morphological parameters and photometry.}\label{tab:molphology}
\begin{threeparttable}[t]
    \centering
    \begin{tabular}{llllll}
    \hline
    Filter & $\lambda_{\rm pivot}/{\rm \mu m}$\tnote{a} & $r_e$/kpc & $n$ & $m_{\rm S\acute{e}rsic}$ & $m_{\rm PSF}$ \\
    \hline
    F606W  & 0.589 & $2.0\pm0.3$ & $2.0\pm0.8$ & $25.9\pm0.2$ & $27.9\pm0.2$ \\
    F814W & 0.804 & $1.6\pm0.4$ & $0.9\pm0.6$ & $25.4\pm0.1$ & $27.2\pm0.2$ \\
    F090W & 0.903 & $1.7\pm0.4$ & $1.0\pm0.7$ & $25.0\pm0.3$ & $26.2\pm0.1$ \\
    F115W & 1.154 & $2.5\pm0.1$ & $1.4\pm0.1$ & $23.4\pm0.1$ & $24.7\pm0.2$ \\
    F150W & 1.501 & $2.94\pm0.07$ & $1.4\pm0.1$ & $22.29\pm0.06$ & $23.23\pm0.04$ \\
    F200W & 1.988 & $3.05\pm0.03$ & $1.60\pm0.04$ & $21.66\pm0.02$ & $22.23\pm0.01$ \\
    F277W & 2.776 & $2.9\pm0.1$ & $1.3\pm0.1$ & $21.3\pm0.1$ & $21.67\pm0.09$ \\
    F356W & 3.566 & $2.9\pm0.4$ & $1.6\pm0.6$ & $21.0\pm0.3$ & $21.1\pm0.4$ \\
    F410M & 4.083 & $2.8\pm0.2$ & $1.3\pm0.7$ & $20.9\pm0.2$ & $20.8\pm0.1$ \\
    F444W & 4.401 & $2.2\pm0.2$ & $2.0\pm1.1$ & $20.6\pm0.1$ & $20.7\pm0.2$ \\
    F770W & 7.639 & $1.7\pm0.2$ & $1.9\pm1.4$ & $21.0\pm0.4$ & $19.3\pm0.3$ \\
    F1800W\tnote{b} & 17.984 & - & - & - & $18.2\pm0.3$ \\\hline
    \end{tabular}
    \begin{tablenotes}
    \item[a] Pivot wavelength $\lambda_{\rm pivot}$ is defined to satisfy $F_\lambda\lambda_{\rm pivot}^2 = F_\nu c$ \citep{Tokunaga2005}, and each value is from \href{https://hst-docs.stsci.edu/wfc3ihb/chapter-6-uvis-imaging-with-wfc3/6-5-uvis-spectral-elements}{HST User documentation} and \href{https://jwst-docs.stsci.edu/jwst-near-infrared-camera/nircam-instrumentation/nircam-filters#gsc.tab=0}{JWST User Documentation}.
    \item[b] Disk (Sérsic component) is undetected based on a comparison of BIC values between the disk + PFS model and the PSF model.
    \end{tablenotes}
\end{threeparttable}
\end{table}

\subsection{Morphology of the host galaxy}\label{subsec:hostmorph}
We detect a disk component in all filters with $n\sim1-2$ and $r_e\sim{\rm 3~kpc}$.
Note that the rest-UV filters, i.e., F606W - F090W, show a compact morphology compared to the rest-optical filters.
In Appendix~\ref{app:deom_full}, we discuss this feature in detail.

\cite{Belli:2014:ApJL} have reported $n\sim5.4$ from an HST/ACS F160W image, but our results indicate the presence of a disk component in the JWST/NIRCam F150W image that has a similar wavelength coverage with HST/ACS F160W.
This discrepancy arises because \cite{Belli:2014:ApJL} have utilized a single Sérsic model without a PSF component, whereas our study employs a composite model of PSF+Sérsic components. 
Indeed, fitting the JWST/NIRcam F150W image with a single Sérsic component gives $n\sim 7$, i.e., hitting the upper limit.

We also confirm that the Bayesian Information Criterion (BIC) for the single Sérsic or the single PSF model is significantly higher than for the PSF + single Sérsic model in all filters except for F1800W. 
These results strongly support that COS-XQG1 is composed of a compact PSF component and an extended disk component (we discuss the nature of the compact PSF component in Section~\ref{subsec:compact}).
These findings, consistent with \cite{Ito2024}, suggest a higher disk fraction in higher-$z$ quiescent galaxies, which might suggest that high-$z$ quiescent galaxies are in a different evolutionary phase from the typical elliptical quiescent galaxies in the local Universe.

\begin{figure}
    \includegraphics[width=8.5cm]{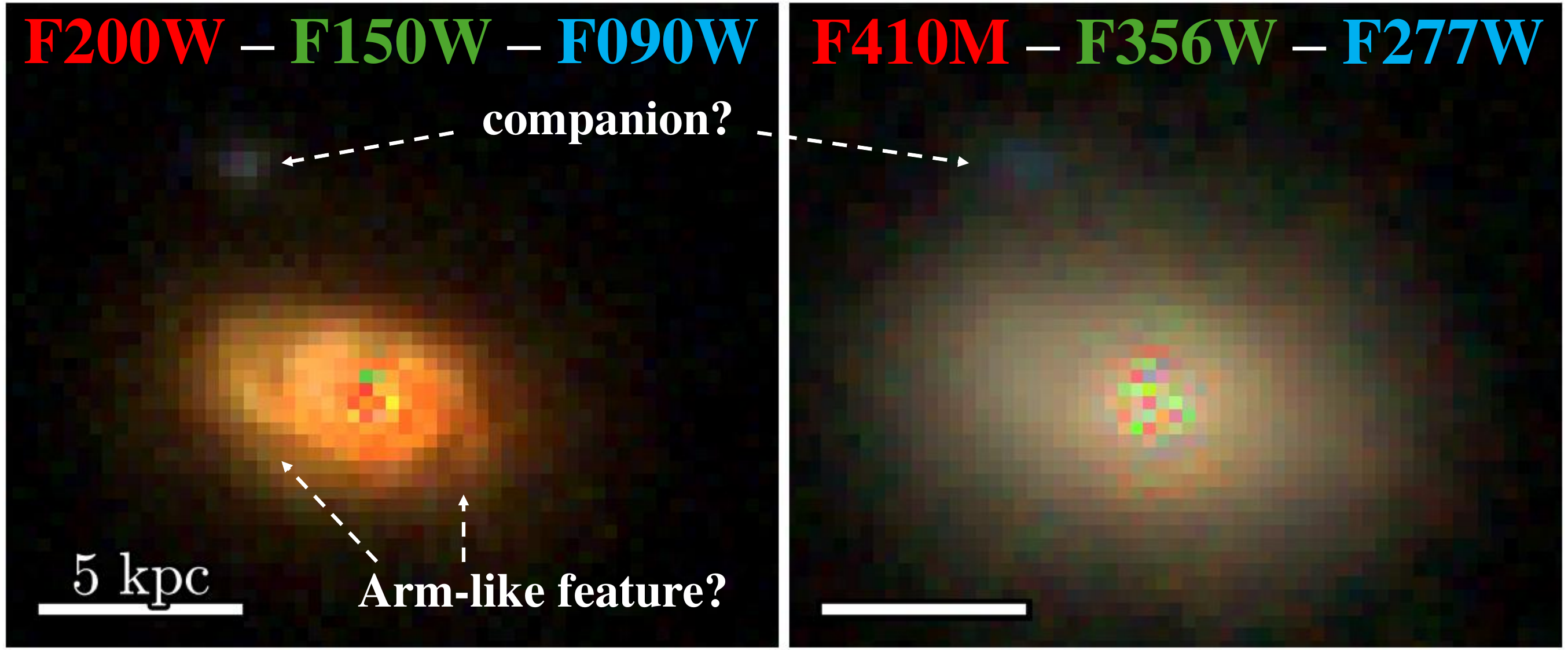}
    \caption{RGB three-color PSF-subtracted images of COS-XQG1.
    R, G, and B colors correspond to F277W, F150W, and F115W (left panel) and to F410M, F356W, and F277W (right). The size of each image is $2\farcs4\times2\farcs0$, and the lower left white bars in individual panels indicate 5~kpc (corresponding to $\sim0\farcs6$).
    }
    \label{fig:3col}
\end{figure}

\begin{figure*}
    \includegraphics[width=17cm]{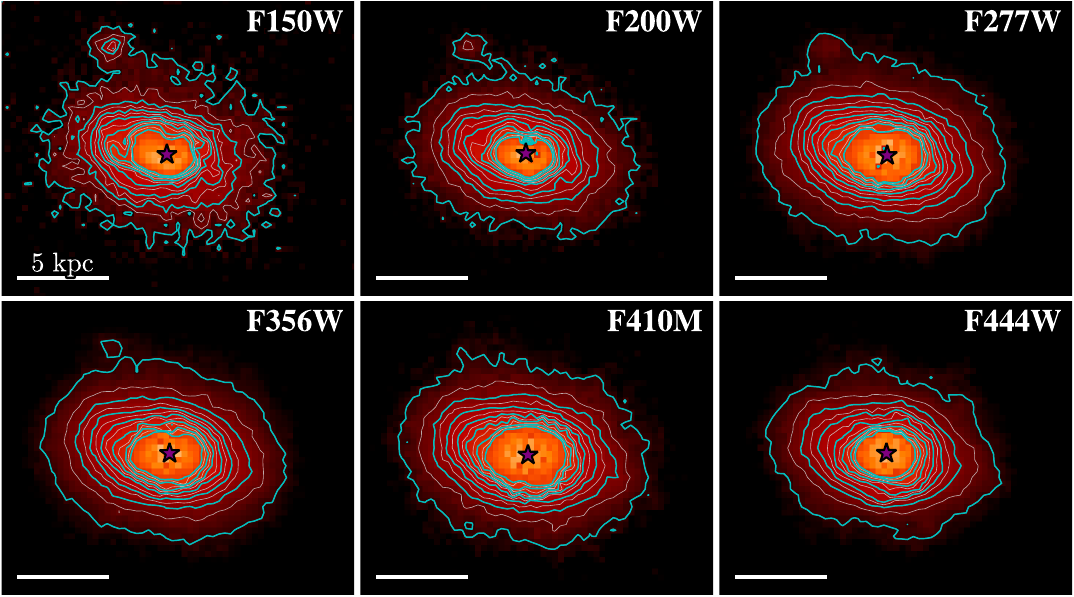}
    \caption{PSF subtracted images in six filters (F150W - F444W) with contours.
    For each image, contour levels are shown from 2\% to 26\% of the peak surface brightness at intervals of $2\%$, with every $4\%$ interval contour indicated with a cyan line. The size of each image is $2\farcs4\times2\farcs0$, and the lower left white bars in individual panels indicate 5~kpc (corresponding to $\sim0\farcs6$).}
    \label{fig:cont}
\end{figure*}

Figure~\ref{fig:3col} displays JWST 3-color PSF-subtracted images from both short (F200W-F150W-F090W) and long (F410M-F356W-F277W) wavelengths.
Figure~\ref{fig:cont} shows isophotal maps of PSF-subtracted images in six filters.
As seen in both figures, short-wavelength images reveal arm-like features extending eastward and westward from the center of the disk component. 

Correspondingly, we can see an asymmetric disk component in long-wavelength contours, i.e., a distorted disk component, especially on the east side.
To assess these features quantitatively, we fit the contours in each PSF-subtracted image with ellipses and find the fitted position angle (PA) changes by $\sim 10$ degrees between the inner ($r\sim2~{\rm kpc}$) and outer regions ($r\gtrsim4~{\rm kpc}$).
Additionally, while the $y$-coordinates of the ellipse centers remain nearly constant, the $x$-coordinates shift eastward.
These results suggest that the disk component is distorted and extends eastward in the outer region.
Such asymmetric substructures, possibly corresponding to an asymmetry of arm structures, are caused by interactions between galaxies (e.g., \citealt{DePropris2007, Bottrell2024}), implying that COS-XQG1 might be experiencing or have experienced an interaction or minor merger with a nearby galaxy.

Furthermore, in the wide wavelength range from HST/ACS F606W to JWST/NIRCam F356W, we identify a nearby component northeast of the COS-XQG1 disk.
This nearby component is fitted as another Sérsic profile.
Assuming the same redshift as COS-XQG1 obtained from JWST/NIRSpec spectrum (Section~\ref{subsec:ppxf}), this nearby object has $n\sim1.3$ and $r_e\sim0\farcs09$ ($\sim 0.7~{\rm kpc}$) in F090W, corresponding to $\sim 3$ times larger than the F090W PSF FWHM ($\sim0\farcs03$).
The separation of 7~kpc from the center of the fitted PSF component of COS-XQG1 suggests that this object could be a counterpart of the aforementioned merger or interaction.
However, we cannot obtain a reliable photo-$z$ estimate for this nearby component due to the large uncertainties in the photometry and difficulties in determining its SED.
Therefore, we cannot conclusively state that this component is a satellite galaxy or a merging counterpart.

\subsection{Nature of the compact component}\label{subsec:compact}

\begin{figure}
    \includegraphics[width=8cm]{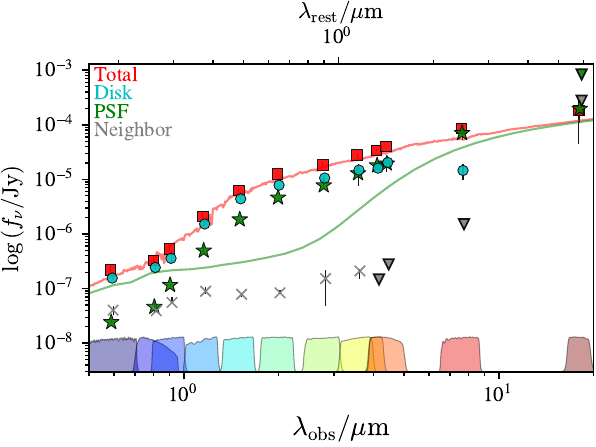}
    \caption{Comparison of the 2D decomposition results and the fitted {\sc XCIGALE} model SED for COS-XQG1.
    Red and green solid lines indicate the model SEDs of total and AGN components, respectively (same as Figure~\ref{fig:CIGALE-SED}). Red squares, cyan circles, green stars, and gray crosses indicate the photometry of the total, disk component, PSF component, and companion object, respectively.
    Downward triangles show $3\sigma$ upper limits in the case of non-detection. The throughput of each filter is shown at the bottom.}
    
    \label{fig:SED_PSF}
\end{figure}

Based on the 2D decomposition analysis (Section~\ref{subsec:decomposition}), we find that COS-XQG1 contains both a compact PSF component and an extended disk component (section~\ref{subsec:hostmorph}).
Given the X-ray detection with a 2--10~keV X-ray luminosity of $\sim9\times10^{43}~{\rm erg~s^{-1}}$, a simple interpretation is that the PSF component comes from AGN emission.
To test this interpretation, we compare the {\sc XCIGALE} model SED of the AGN component (Section~\ref{subsec:CIGALE}) with the photometry of the PSF component from the 2D decomposition (Section~\ref{subsec:decomposition}) in Figure~\ref{fig:SED_PSF}.

The results show a significant discrepancy: the photometry of the PSF component exceeds the AGN model SED, especially in rest-optical to rest-NIR wavelengths ($\sim0.3-2~{\rm \mu m}$). 
In rest-UV wavelengths, on the other hand, the AGN model SED exceeds the SED of the PSF component.
Especially in F277W, while the model SED predicts that the AGN component accounts for only $\sim 5\%$ of the total flux, the 2D decomposition analysis (Section~\ref{subsec:decomposition}) clearly detects a PSF component accounting for $42\pm6\%$ of the total flux.

We propose the presence of a compact bulge as the cause of this discrepancy.
The power-law fitting of the F115W-F200W photometry of the PSF component results in a rest-optical spectral slope of $\beta\sim 2.1\pm0.3$ ($F_\lambda\propto\lambda^\beta$).
This value is significantly different from typical AGN values $\beta\sim-1.5$ (e.g., \citealt{Vanden2001}) and suggests an SED similar to a galaxy component or a dust-obscured AGN.
Although the {\sc XCIGALE} SED fitting also allows for a dust-obscured AGN solution by increasing the extinction of the polar dust, our SED fitting results in a polar torus dust emission with $E\left(B-V\right)\sim0$, suggesting such a dust-obscured AGN model cannot fully explain the PSF component.

Furthermore, as shown in Figure~\ref{fig:SED_PSF}, the SED of the PSF component has a spectral slope of $\beta\sim3.6$ at wavelengths shorter than rest-4000~\AA\ but $\beta\sim0.1$ at wavelengths longer than rest-4000~\AA.
This slope change is consistent with the Balmer break in the PSF component, i.e., the presence of a stellar component.
Therefore, it is likely that COS-XQG1 has a compact stellar component, possibly a bulge, that is difficult to separate from the AGN component in the 2D decomposition analysis.
We attempt 2D decomposition fitting with a PSF and double Sérsic components to further separate the bulge and the AGN.
However, possibly due to the degeneracy between the compact bulge and the AGN, we cannot obtain consistent results across different bands.
This discrepancy suggests the limitation of the current 2D decomposition method, demonstrating the necessity for considering the possibility of contamination by the central compact stellar component in a similar 2D decomposition analysis and for improving the current decomposition techniques.
One possible solution is to resolve the degeneracy that cannot be solved with imaging alone due to PSF effects by inputting the SED information into the image-based decomposition \citep[e.g.,][]{Yu2024, Chen2024}.

Also, note that the additional stellar components cannot describe the excess of the AGN model SED in rest-UV wavelengths.
Because the AGN is not spatially resolved by JWST, AGN emission is expected to originate from a compact central region at the same position across all filters, i.e., the PSF component, whose position is almost fixed in the 2D decomposition analysis (Section~\ref{subsec:decomposition}).
Therefore, PSF components should at least include the AGN component and possibly include compact stellar components if the galaxy has such a compact bulge.
For this reason, the model SED overestimates the AGN components in rest-UV, indicating the need for modifying SED fitting results.
To obtain consistent SED fitting results, we attempt to use strong constraints on the parameter grid values used in the SED fitting.
However, these SED fittings result in poor fits in other photometry and an increase in the $\chi^2$ value.
We also confirm that the estimated $M_*$ and SFR under the strong constraints remain consistent with the values estimated under the weak constraints (Section~\ref{subsec:CIGALE}) within the errors.
Therefore, in this paper, we use the results with the smallest $\chi^2$ value under the weak constraints (Section~\ref{subsec:CIGALE} and Figure~\ref{fig:CIGALE-SED}).
This result also indicates the need for future improvements in the arts of SED fitting, such as improvements in template SEDs.
Parallel fitting of 2D decomposition and the SED fitting discussed above might also enable the inclusion of image-based information in the SED fitting, possibly solving the issue.

\section{Discussion}
\subsection{Connection between AGN and quenching}\label{subsec:propCOS-XQG1}
\begin{figure}
    \includegraphics[width=8.5cm]{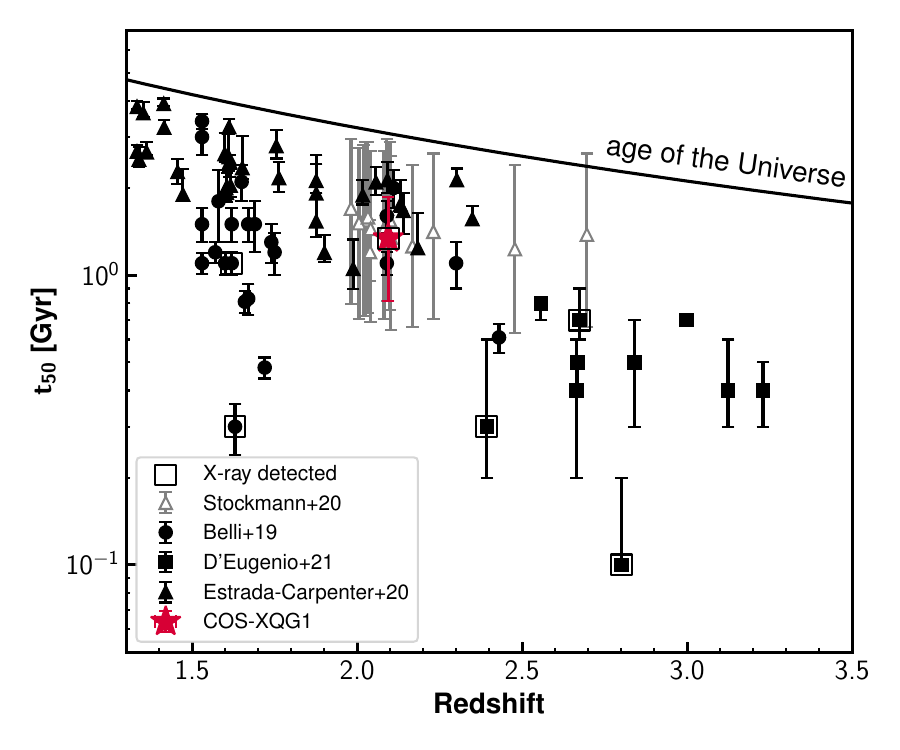}
    \caption{Stellar age of quiescent galaxies as a function of redshift. The look-back time after 50\% of the stellar mass was formed (i.e., $t_{\rm 50}$) of COS-XQG1 is shown as a red star. Those of other quiescent galaxies at $1.5<z<3.5$ are shown as black markers \citep{Belli:2019:ApJ, Estrada-Carpenter:2020:ApJ, D'Eugenio:2021:A&A}. The mass-weighted ages of quiescent galaxies from \citet{Stockmann:2020:ApJ} are shown as gray triangles. Those detected in the X-ray are enclosed by a black square. A solid line shows the age of the Universe as a function of redshift.}
    \label{fig:t50}
\end{figure}

In Section \ref{subsec:CIGALE}, we have found that COS-XQG1 formed its half of the stellar mass $\sim1.3$ Gyrs ago. We compare this age with those of other quiescent galaxies at a similar redshift. Figure \ref{fig:t50} shows the stellar age of quiescent galaxies as a function of redshift. We here refer to the studies which derive $t_{\rm 50}$ \citep{Belli:2019:ApJ, Estrada-Carpenter:2020:ApJ, D'Eugenio:2021:A&A}. COS-XQG1 is slightly younger than the other known quiescent galaxies at a similar redshift because most of the quiescent galaxies at $2\leq z \leq 2.2$ have $t_{\rm 50} \sim2\, {\rm Gyr}$ \citep[][]{Belli:2019:ApJ, Estrada-Carpenter:2020:ApJ}. Other X-ray-detected quiescent galaxies seem to follow this trend. This might imply that quiescent galaxies with an active AGN activity are more recently formed than those without. If this is the case, an observed AGN activity of a quiescent galaxy is related to its quenching even though quenching itself happened well before the observed epoch.

Despite being a quiescent galaxy, COS-XQG1 has a clear disk-like morphology, as shown in Figure~\ref{fig:decom}.
Such not-star-forming spiral or disk galaxies, which are called passive spirals, red spirals, or anemic spirals, have been found in the local Universe (e.g., \citealt{vandenBergh1976, Shimakawa2022}) and in the high-$z$ Universe (e.g., \citealt{Fudamoto2022}).
Red spiral galaxies have been considered as spiral galaxies that have quenched due to AGN feedback (e.g., \citealt{Lokas2022}).
Therefore, COS-XQG1 might also have been quenched due to the AGN activity that lasted.

\subsection{Black hole-galaxy co-evolution at $z\sim2$}
\begin{figure*}
    \includegraphics[width=17cm]{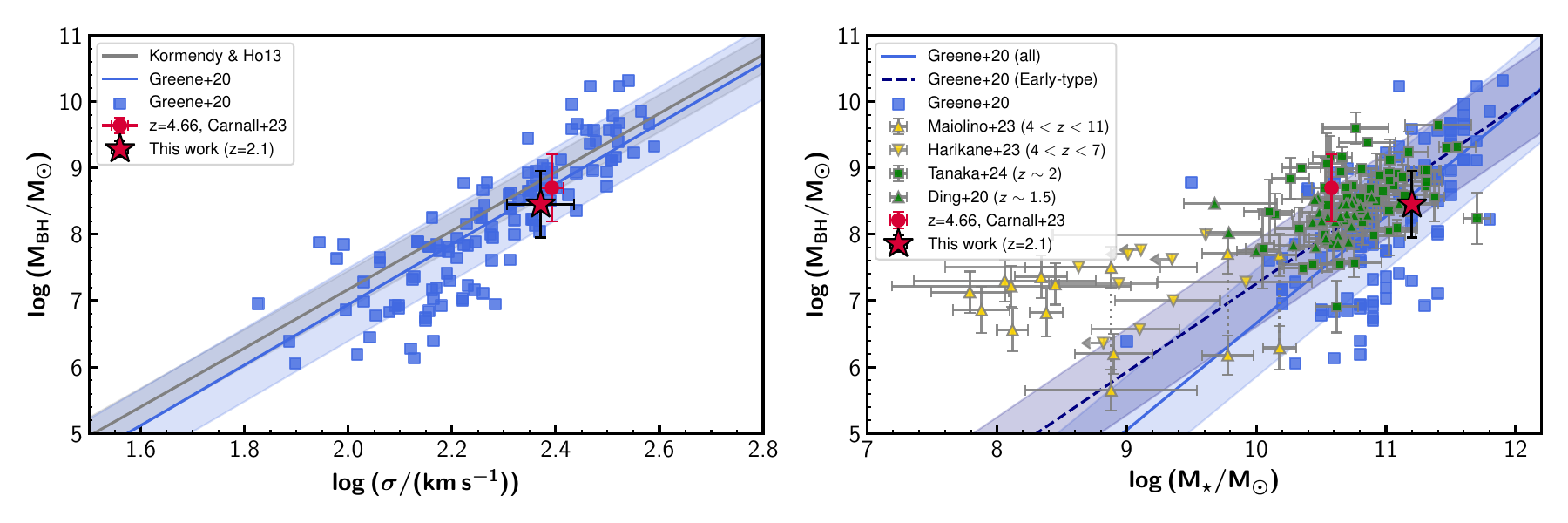}
    \caption{Left panel: The black hole mass -- stellar velocity dispersion relation. COS-XQG1 is shown as a red star. The systematic uncertainty in the black hole mass measurement ($\sim0.5$ dex) is included in its error bar. The local scaling relations from \citet{Greene:2020:ARA&A} and \citet{Kormendy2013} are shown as solid blue and gray lines, respectively. The hatched region of each color corresponds to the range of the intrinsic scatter of each relation. A red circle corresponds to a quiescent galaxy at $z=4.66$ from \citet{Carnall:2023:Natur}, where the error in $M_{\rm BH}$ includes the systematic uncertainty as in the case of COS-XQG1. Blue squares correspond to local objects compiled in \citet{Greene:2020:ARA&A}. Right panel: The black hole mass -- stellar mass relation. The meaning of symbols is the same as the left panel. The dashed line is the local scaling relation from \citet{Greene:2020:ARA&A} for only elliptical galaxies. Individual measurements of objects at a wide redshift range are also shown. Green squares and triangles are galaxies at $z\sim2$ and $z\sim1.5$ from \citet{Tanaka2024} and \citet{Ding2020_HST}, respectively. Yellow markers corresponds to objects at higher redshift from \citet{Maiolino:2023:arXiv} (triangles, $4<z<11$) and \citet{Harikane:2023:ApJ} (inverted triangles, $4<z<7$). }
    \label{fig:BHgalcoevl}
\end{figure*}

Using the stellar velocity dispersion measurement (Section \ref{subsec:ppxf}) and the black hole mass measurement (Section \ref{subsec:bhmass}), we plot COS-XQG1 in the $M_{\rm BH} - \sigma_\star$ plane in Figure \ref{fig:BHgalcoevl}. 
This galaxy is the second one for $z\geq2$, following GS-9209 \citep{Carnall:2023:Natur}, that has both a black hole mass measurement and a direct measurement of stellar velocity dispersion as far as we know \citep[see ][
which have used gas velocity dispersion]{Ubler:2023:A&A, Maiolino:2023:arXiv}. COS-XQG1 follows the relation in the local Universe \citep[][]{Kormendy2013, Greene:2020:ARA&A}. GS-9209 also follows the local relation. Although the number of sources should be increased in the future, these two galaxies might suggest that high-$z$ massive quiescent galaxies, which are likely the progenitor of local elliptical galaxies, form the same $M_{\rm BH} - \sigma_\star$ relation as in the local Universe as early as $z\geq2$. 

We then plot COS-XQG1 in the $M_{\rm BH} - M_\star$ plane using the stellar mass from {\sc XCIGALE} fitting (Section \ref{subsec:CIGALE}), finding that COS-XQG1 is consistent also with the local $M_{\rm BH} - M_\star$ relation \citep{Greene:2020:ARA&A}. The black hole to stellar mass ratio of COS-XQG1 is $M_{\rm BH}/M_\star \sim1.8\times10^{-3}$. This galaxy is located in the massive end of X-ray selected AGNs at $z\sim1.5-2$ \citep{Tanaka2024, Ding2020_HST}, for which the stellar mass of host galaxies has been derived using the 2D image decomposition technique. Again, GS-9209 has a similar trend to COS-XQG1. 

COS-XQG1 and GS-9209 might suggest that quiescent galaxies at $z\geq2$ are generally consistent with the local $M_{\rm BH} - \sigma_\star$ and $M_{\rm BH} - M_\star$ relations in contrast to recently found low-mass AGNs at higher redshift \citep[e.g.,][]{Harikane:2023:ApJ, Maiolino:2023:arXiv}, many of which have over massive black holes compared to the local $M_{\rm BH} - M_\star$ relation ($M_{\rm BH}/M_\odot \sim10^{-2}-10^{-1}$). This difference, if real, could be because quiescent galaxies, including COS-XQG1, are more evolved than other low-mass AGNs. Quiescent galaxies are massive enough and do not have significant star formation, at least at the observed epoch. If they keep the observed low star formation rates, they will not increase the stellar mass significantly \citep[see, however,  discussion on rejuvenation, e.g.,][]{Tanaka:2024:PASJ}. The black hole accretion rate measurement of ${\rm BHAR} \sim 0.22\, M_\odot {\rm yr^{-1}}$ suggests that COS-XQG1's black hole mass also will not increase significantly. It requires $\sim2\, {\rm Gyr}$ to double the mass even if it keeps the current BHAR. Thus, if quiescent galaxies are on the local relation at the observed time, they would also be so after that. On the other hand, most low-mass AGNs are likely star-forming because the fraction of quiescent galaxies decreases with decreasing stellar mass and with increasing redshift \citep[e.g., $\sim 2\%$ at $10<\log{(M_\star/M_\odot)}\leq10.5$ and $z\sim4$][]{Weaver:2023:A&A}. Thus, it is likely that they will grow the stellar mass more significantly and might reach closer to the local relation in later epochs. 

\subsection{Comparison with the Illustris-TNG simulation}

We have found a quiescent galaxy at $z=2.09$ that lies on the black hole - galaxy scaling relation in the local Universe. 
We compare quiescent galaxies in the IllustrisTNG simulation \citep{Nelson:2019:ComAC} and discuss how quiescent galaxies at high redshift have evolved until $z=2.1$ both in terms of stellar and black hole masses.

We first select quiescent galaxies at $z=2.10$ (Snapshot 32) from the TNG-300, the largest box simulation among the current TNG runs, by imposing an upper limit on the specific star formation rate as $\log{\rm (sSFR/yr^{-1})}\leq-10$. This specific star formation rate is measured within twice the stellar half-mass radius, as done in our previous works \citep{Valentino:2020:ApJ, Ito:2023:ApJL}. Here, a star formation rate averaged over 10 Myrs is used. Furthermore, we impose a stellar mass cut of $M_\star\geq 1.5\times 10^{11}\, M_\odot$ to only focus on a similar stellar mass range to COS-XQG1. As a result, we select 15 quiescent galaxies at $z=2.1$.

\begin{figure*}
    \includegraphics[width=17cm]{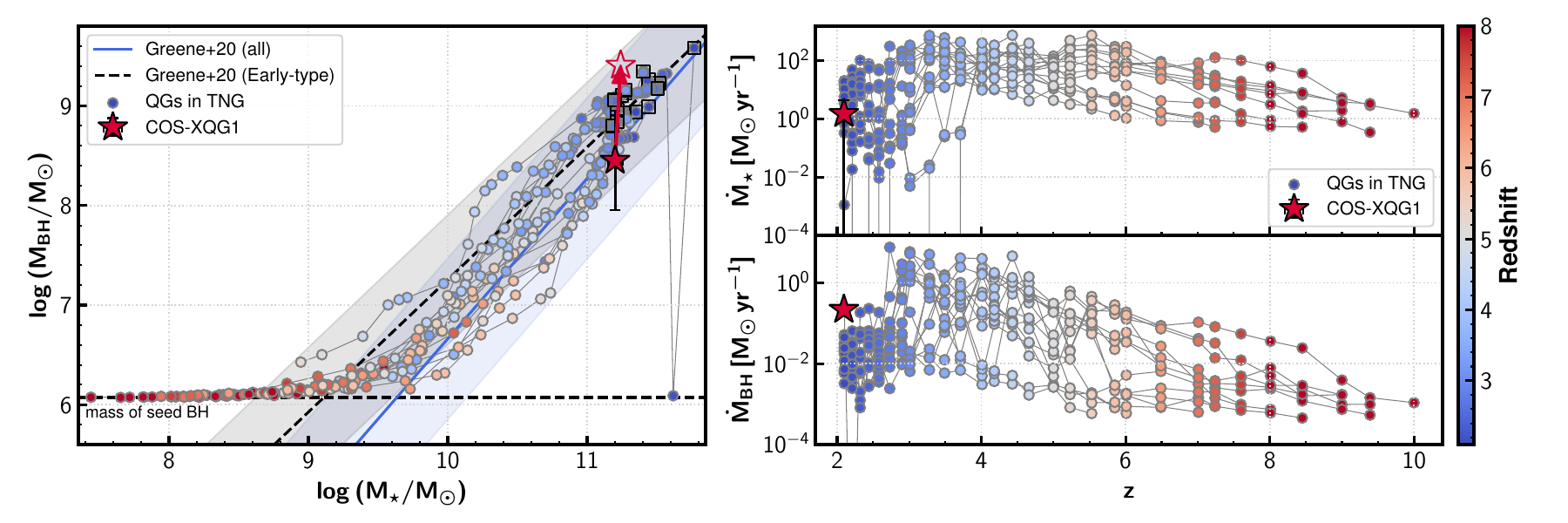}
    \caption{Left panel: The evolution of quiescent galaxies (circles) in TNG-300 in the $M_{\rm BH} - M_\star$ plane
    color-coded by redshift, with positions at $z=2.1$ highlighted with black squares. COS-XQG1 is shown as a red star. Assuming that the $\dot{M_\star}$ and $\dot{M_{\rm BH}}$ are constant from $z=2.0943$ to $z=0$, the growth vector of the stellar mass and black hole mass growth from $z=2.0943$ to $z=0$ and its position at $z=0$ of this case are shown as a red arrow and a red open star, respectively. Also plotted are observationally determined local relations for all galaxies (solid line) and elliptical galaxies (dashed line) given in \citet{Greene:2020:ARA&A}. Right panel: The redshift evolution of star formation rate (upper panel) and black hole accretion rate (bottom panel) of quiescent galaxies in TNG-300. The meaning of symbols is the same as the left panel.}
    \label{fig:TNGcomp}
\end{figure*}

Figure \ref{fig:TNGcomp} shows the evolution of these quiescent galaxies in TNG-300 in the $M_{\rm BH} - M_\star$ and $\dot{M}_{\rm BH} - \dot{M}_\star$ planes from $z\sim8$ to $z=2.1$. Firstly, we focus on the $M_{\rm BH} - M_\star$ relation. At $z=2.1$, the quiescent galaxies in TNG-300 are consistent with the local relation \citep[e.g.,][]{Greene:2020:ARA&A}, similar to COS-XQG1. At higher redshift (e.g., $z\geq4$), they are still consistent with the relation of all local galaxies \citep[e.g.,][]{Greene:2020:ARA&A} but below the relation of elliptical galaxies. This trend is reasonable since most have not yet been quenched at that epoch. At least, quiescent galaxies selected here do not have a significant offset from the local observationally determined relation. In addition, they do not have over-massive black holes throughout their history. A similar trend is reported in a different simulation, such as the Magneticum Pathfinder simulation \citep[][]{Kimmig:2023:arXiv}.

The redshift evolution of star formation rate and black hole accretion rate (the right panel of Figure \ref{fig:TNGcomp}) shows that the quiescent galaxies have a peak star formation rate and a peak black hole accretion rate at almost the same time, typically at $z\sim 4$. At redshift $z=2.1$, they have ${\rm BHAR}\sim10^{-2}\, M_\odot {\rm yr^{-1}}$. This value is in agreement with an observational constraint on the average black hole accretion rate derived from an average X-ray luminosity at $2<z<2.5$ \citep[${\rm BHAR}=7\times10^{-3}\, M_\odot {\rm yr^{-1}}$,][]{Ito:2022:ApJ}. COS-XQG1 has a higher BHAR than the TNG-300 quiescent galaxies, likely because this object is selected as an X-ray-detected galaxy.

Suppose that COS-XQG1 follows the typical evolutional track of quiescent galaxies in TNG-300. In that case, it is likely that at the observed time ($z=2.09$), COS-XQG1 has ended its stellar mass growth but still has a residual black hole accretion of past activity. This galaxy might have been quenched more recently than typical quiescent galaxies and still has a higher AGN activity. The younger age of the COS-XQG1 than the other observed quiescent galaxies at the same redshift (Section \ref{subsec:propCOS-XQG1}) is in line with this possible scenario. If so, its black hole accretion rate will decrease and reach the value of the typical quiescent galaxies in TNG later. We have to note that this is just a case study. It is necessary to examine this trend seen in TNG-300 using an increased number of quiescent galaxies with black hole mass and black hole accretion rate measurements.

\section{Conclusion}\label{sec:concl}
In this paper, we report a characterization of an X-ray-detected quiescent galaxy at $z=2.09$, COS-XQG1. This galaxy was initially reported in \citet{Belli:2014:ApJL} and recently observed with NIRSpec, NIRCam, and MIRI on JWST.

We first conducted SED fitting analysis combining multi-wavelength photometry from X-ray to submillimeter, including the NIRCam and MIRI photometry. We have confirmed that this galaxy is indeed a massive ($M_\star = (1.6 \pm 0.2) \times 10^{11}\, M_\odot$) quiescent (${\rm sSFR} < 10^{-10}\, {\rm yr^{-1}}$) galaxy even we consider a possible contribution of the AGN component to the SED. The stellar component dominates at $\lambda > 1\, {\rm \mu m}$, suggesting that we can constrain its host galaxy property.

COS-XQG1 has several broad emission lines, e.g., H$\beta$ and H$\alpha$, which is typical of Type 1 AGNs. We have modeled the H$\alpha$ emission and measured the FWHM and flux of the broad component. The FWHM is as broad as ${\rm FWHM_{H\alpha}}=4365^{+81}_{-81}\, {\rm km\, s^{-1}}$, suggesting that this comes from the broad line region of an AGN. Based on them, the black hole mass of COS-XQG1 is estimated as $\log{(M_{\rm BH}/M_\odot)} = 8.43\pm0.02(\pm 0.5)$. Using the stellar velocity dispersion measured from stellar absorption lines in the NIRSpec spectrum, we have put the second data point of $z\geq2$ in the $M_{\rm BH} - \sigma_\star$ relation in addition to a quiescent galaxy at $z=4.6$ from \citet{Carnall:2023:Natur}. We do not see a strong deviation of these data points at $z\geq 2$ from the local relation \citep[e.g.,][]{Kormendy2013, Greene:2020:ARA&A}. The same trend is also seen in the $M_{\rm BH} - M_\star$ relation. This result may suggest that massive quiescent galaxies have already matured in terms of both stellar and black hole masses. Indeed, observations of high-$z$ quiescent galaxies have found that they have only a small amount of gas \citep[e.g.,][]{Suzuki:2022:ApJ, Caliendo:2021:ApJL}.

From our 2D decomposition analysis, we have detected both an extended Sérsic component and a PSF component at the center.
The Sérsic component has an extended disk component with 
$r_e \sim 3~{\rm kpc}$ and $n\sim 1.5$.
The disk also exhibits substructures such as arms and distortions, suggesting the possibility of recent interactions.
We have also found a discrepancy between the SED of the PSF component and the AGN model SED predicted by the SED fitting using total photometry.
The SED of the PSF component shows a feature resembling the Balmer break, suggesting the need to modify the current 2D decomposition analysis method.

We have compared COS-XQG1 with quiescent galaxies at the same redshift in the Illustris TNG-300 simulation. Quiescent galaxies at $z=2.1$ in TNG-300 are also located in the local $M_{\rm BH} - M_\star$ relation, similar to COS-XQG1. The black hole accretion rate of COS-XQG1 is slightly higher than that of the quiescent galaxies in TNG-300, probably because we selected this galaxy as an X-ray source. However, this galaxy is on the evolution track of quiescent galaxies in TNG-300, implying that if the evolution of quiescent galaxies in TNG-300 is real, this galaxy is on the way to getting quenched in terms of both stellar and black hole mass growth.

This study demonstrates the potential usefulness of X-ray-detected quiescent galaxies in understanding the co-evolution between black holes and galaxies and the role of AGN in galaxy quenching. Increasing the spectroscopic sample of this galaxy population, especially by observing stellar absorption lines and H$\alpha$ emission, is essential to obtain a census of the $M_{\rm BH} - \sigma_\star$ relation at high redshift.

\section*{Acknowledgements}
This study was supported by JSPS KAKENHI Grant Numbers JP22J00495 and JP23K13141. This study is based on the observations associated with JWST programs GO \#1810 and \#1837. The authors acknowledge the teams and PIs for developing their observing program. The data products presented herein were retrieved from the Dawn JWST Archive (DJA). DJA is an initiative of the Cosmic Dawn Center (DAWN), which is funded by the Danish National Research Foundation under grant DNRF140. 
Numerical computations were in part carried out on the iDark cluster, Kavli IPMU.
Kavli IPMU is supported by World Premier International Research Center Initiative (WPI), MEXT, Japan.
TT is supported by the Forefront Physics and Mathematics Program to Drive Transformation (FoPM), which is a World-leading Innovative Graduate Study (WINGS) Program of the University of Tokyo.

\section*{Data Availability}
The images and spectrum used in this study are taken from the DAWN {\it JWST} Archive (DJA, \url{https://dawn-cph.github.io/dja/}). Data directly related to the figures of this publication are available upon request to the corresponding author.



\bibliographystyle{mnras}
\bibliography{ito,tanaka} 




\appendix
\section{Comparison of the spectra of the broad H$\alpha$ line taken with Keck/MOSFIRE and JWST/NIRspec} \label{sec:MOSFIREHalpha}

\begin{figure}
    \centering
    \includegraphics[width=8.5cm]{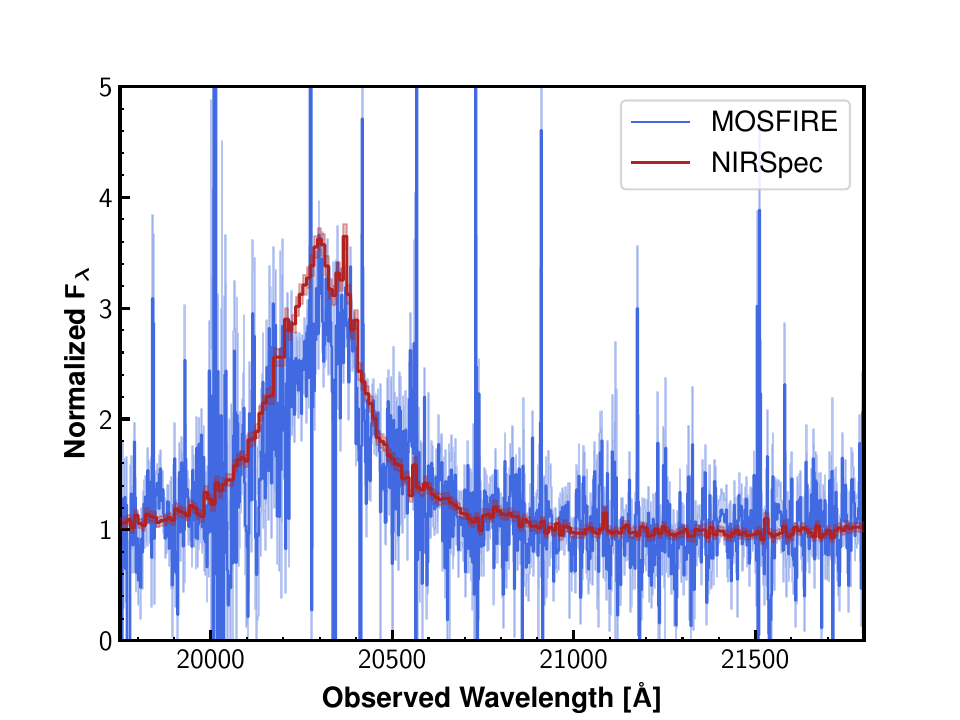}
    \caption{Normalized spectra around the H$\alpha$ emission line of COS-XQG1 obtained with Keck/MOSFIRE (blue) and JWST/NIRSpec (red). The uncertainty in each spectrum is shown as a shaded region with the same color.}
    \label{fig:MOSFIRE-JWST-Halpha}
\end{figure}

In figure \ref{fig:MOSFIRE-JWST-Halpha}, we show the spectra around the H$\alpha$ emission line of COS-XQG1 obtained with Keck/MOSFIRE and JWST/NIRSpec. The MOSFIRE spectrum is taken from the public data release of the ZFIRE survey \citep{Nanayakkara:2016:ApJ}. 
The exact amplitudes of the two spectra are different because of different slit losses and different flux correction methods. Thus, we here show normalized spectra. It is evident that the broad line shape is consistent between the two spectra and that the JWST/NIRSpec spectrum has much higher signal-to-noise ratios.

\section{Posterior distribution of parameters in the line fitting} \label{app:conerplot}

\begin{figure*}
    \centering
    \includegraphics[width=1\linewidth]{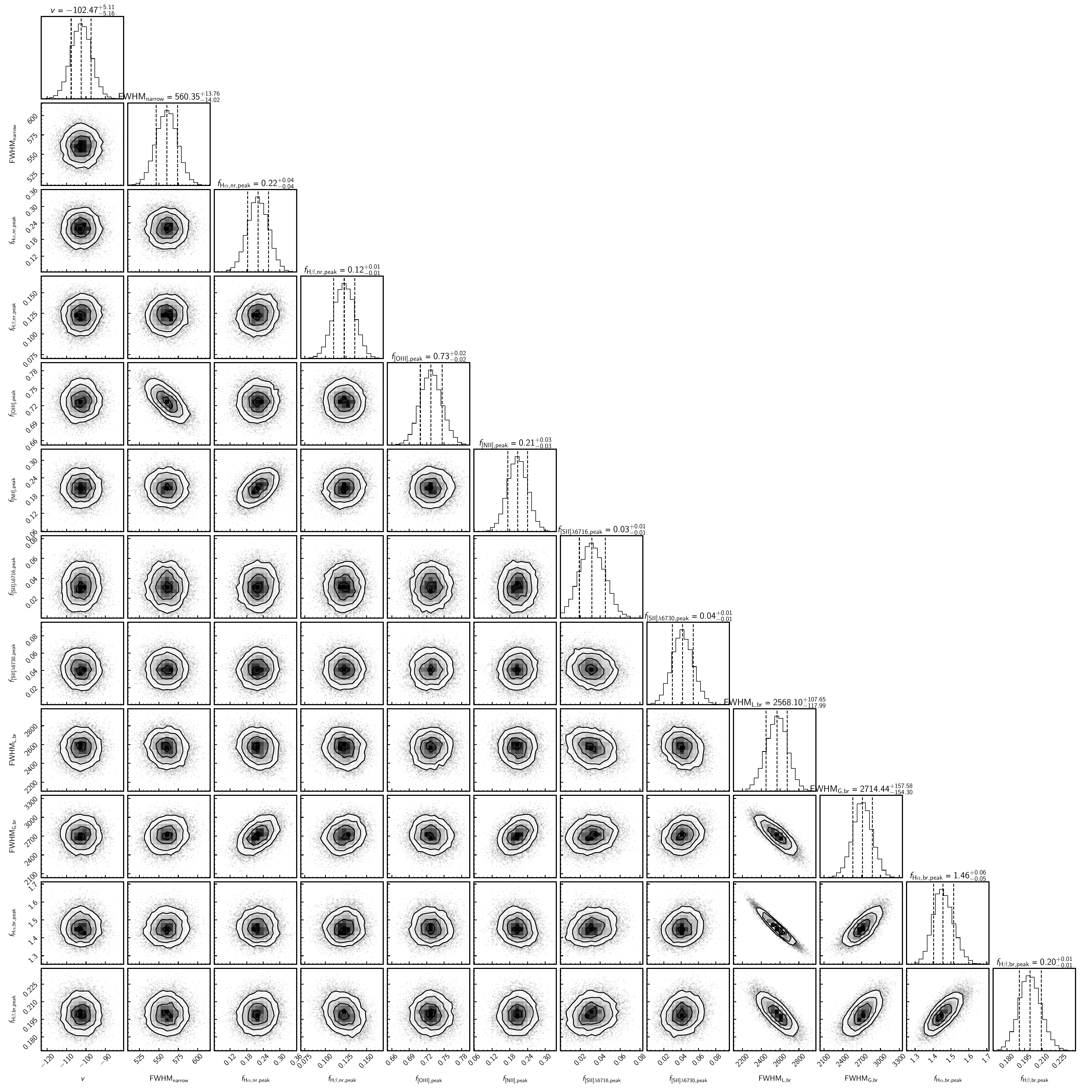}
    \caption{The posterior distribution of parameters in the line fitting. The units of all parameters are arbitrary.}
    \label{fig:corner}
\end{figure*}

Figure \ref{fig:corner} shows the posterior distribution of parameters in the line fitting conducted in Section \ref{subsec:halpha}.

\section{Discussion of the full 2D Decomposition results}\label{app:deom_full}
\begin{figure*}
    \includegraphics[width=17cm]{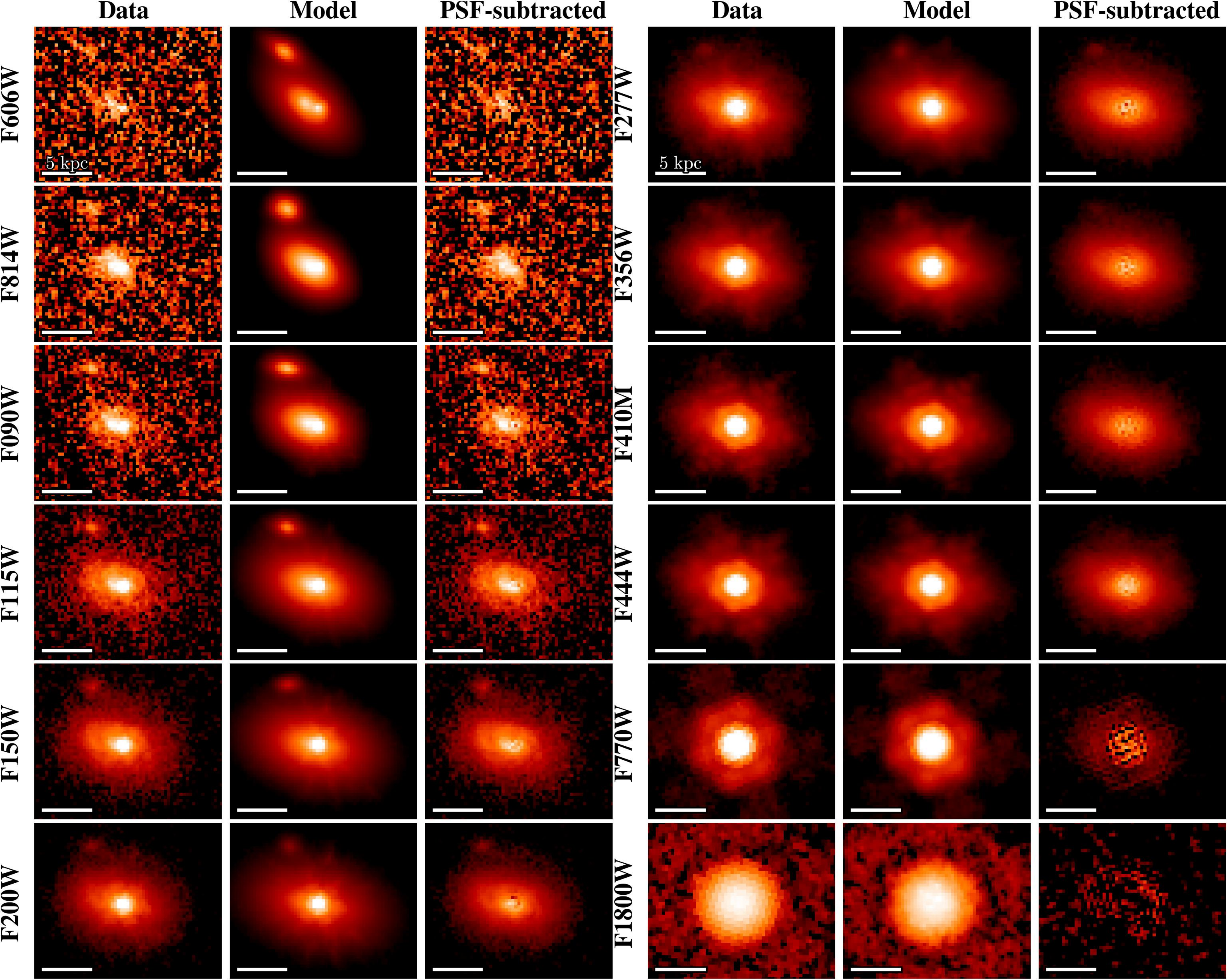}
    \caption{2D decomposition fitting in each filter (full filter version of Figure~\ref{fig:decom}). The size of each image is $2\farcs4\times2\farcs0$, and the lower left white bars in individual panels indicate 5~kpc (corresponding to $\sim0\farcs6$).
    }
    \label{fig:decom_all}
\end{figure*}

The results of the 2D decomposition analysis (Section~\ref{subsec:decomposition}) for all available filters of HST and JWST (NIRCam and MIRI) are shown in Figure~\ref{fig:decom_all}, which is the full version of Figure~\ref{fig:decom}.
In the JWST/NIRCam bands, the compact component visible in the original image (left panel in each filter row in Figure~\ref{fig:decom_all}) is clearly subtracted in the original $-$ PSF image (right panel in each filter row in Figure~\ref{fig:decom_all}).
Even in the NIRCam long-wavelength bands (F277W - F444W), we clearly detect the disk component that is buried in the PSF component in the original image.

In Section~\ref{subsec:hostmorph} and Table~\ref{tab:molphology}, we find that $r_e$ in the rest-UV (F606W - F090W) are smaller than those in the rest-optical to rest-NIR (F115W - F444W).
From Figure~\ref{fig:decom_all}, we can see that this trend is because the disk component in the rest UV filters, especially in F606W and F814W, is less clear. 
As shown in fitted model images (center panel in each filter row in Figure~\ref{fig:decom_all}), 
the center position and position angle of the Sérsic disks in F606W and F814W are slightly shifted from those in the rest-optical and rest-NIR filters.
The direction of the shift in the central position in these rest-UV filters is consistent with the eastward-extended arm structure seen in F115W and F150W, which may indicate the presence of a relatively younger stellar population there.

Previous studies have not used MIRI imaging data for 2D decomposition analysis because of large PSF sizes, e.g., FWHM$\sim0\farcs3$ in F770W and $\sim0\farcs6$ in F1800W.
In this study, using BIC criteria, we successfully detect the host galaxy of a $z\gtrsim 2$ AGN in F770W for the first time, demonstrating that F770W imaging data can be applied to 2D decomposition analysis at $z\sim2$, at least.
However, in F1800W, BIC criteria indicate that a single PSF model is sufficient to describe the original image, i.e., a Sérsic component is not detected.
The non-detection of a Sérsic component is consistent with the {\sc XCIGALE} decomposition results, suggesting that for our object the AGN component accounts for most of the total flux.

\bsp	
\label{lastpage}
\end{document}